\documentclass[twocolumn,english,aps,pra,reprint, superscriptaddress,showpacs,longbibliography,showkeys, nofootinbib]{revtex4-1}
\usepackage{amsmath,amssymb,mathtools,bbm,mathrsfs,bm,braket,color,graphicx,comment,xcolor,dsfont,multirow}
\usepackage{hyperref}
\hypersetup{colorlinks,allcolors=blue}
\usepackage[mathscr]{euscript}
\usepackage{subfigure}
\usepackage{multirow}
\usepackage{enumerate}
\usepackage{siunitx}  
\usepackage{algorithm}
\usepackage{algpseudocode}

\begin{document}

\title{Runtime Reduction in Linear Quantum Charge-Coupled Devices using the Parity Flow Formalism}

\author{Federico Domínguez}
\affiliation{Parity Quantum Computing Germany GmbH, 20095 Hamburg, Germany} 

\author{Michael Fellner}
\affiliation{Parity Quantum Computing GmbH, A-6020 Innsbruck, Austria}
\affiliation{Institute for Theoretical Physics, University of Innsbruck, A-6020 Innsbruck, Austria}

\author{Berend Klaver}
\affiliation{Parity Quantum Computing GmbH, A-6020 Innsbruck, Austria}
\affiliation{Institute for Theoretical Physics, University of Innsbruck, A-6020 Innsbruck, Austria}

\author{Stefan Rombouts}
\affiliation{Parity Quantum Computing Germany GmbH, 20095 Hamburg, Germany} 

\author{Christian Ertler}
\affiliation{Parity Quantum Computing Germany GmbH, 20095 Hamburg, Germany} 

\author{Wolfgang Lechner}
\affiliation{Parity Quantum Computing Germany GmbH, 20095 Hamburg, Germany} 
\affiliation{Parity Quantum Computing GmbH, A-6020 Innsbruck, Austria}
\affiliation{Institute for Theoretical Physics, University of Innsbruck, A-6020 Innsbruck, Austria}

\begin{abstract}
Using the Parity Flow formalism~\cite{Klaver2024}, we show that physical SWAP gates can be eliminated in linear hardware architectures, without increasing the total number of two-qubit operations. This has a significant impact on the execution time of quantum circuits in linear Quantum Charge-Coupled Devices (QCCDs), where SWAP gates are implemented by physically changing the position of the ions. Because SWAP gates are one of the most time-consuming operations in QCCDs, our scheme considerably reduces the runtime  of the quantum Fourier transform and the quantum approximate optimization algorithm on all-to-all spin models, compared to circuits generated with standard compilers (TKET and Qiskit). While increasing the problem size (and therefore the number of qubits) typically demands longer runtimes, which are constrained by coherence time, our runtime reduction enables a significant increase in the number of qubits at a given coherence time. 

\end{abstract}

\date{\today}
\maketitle

\section{Introduction}

Quantum Charge-Coupled Devices (QCCDs) \cite{Kielpinski2002} are a promising architecture for quantum computing, known for achieving high fidelities and offering all-to-all qubit connectivity through transport operations \cite{moses2023race, Bjoern2017blueprint, malinowski2023howtowire, weidt2016trapped, kaushal2020shuttling}. Most of the currently operational QCCDs exhibit linear qubit connectivity, and while these devices are challenging to scale up due to this linear topology, they may still demonstrate quantum advantage without the need for quantum error correction~\cite{ruslan2024evidence}.

The primary limitation of linear QCCDs is their long runtime, mainly dominated by transport and cooling operations. Quantum gates can only be implemented if the qubits involved are located in a specific part of the quantum chip (known as the \textit{gate zone} or \textit{interaction zone}), where high-power laser light is accessible. Transport operations, including (physical) SWAPs and ion shuttles, are necessary to sort the ions and bring them into the gate zone. However, these operations, along with the subsequently required cooling of the ion chain, are typically time-consuming~\cite{monroe2013scaling, pino2021demonstration}. Although these processes do not significantly reduce the quantum state fidelity~\cite{palani2023high, kaufmann2018high, Blakestad2009high}, the accumulation of transport time will, as the system size increases, eventually face the limit of the coherence time $T_2$. Therefore, efficient algorithms for linear systems that reduce transport time are highly beneficial to achieve larger and more competitive computation in QCCDs.

The Parity Flow formalism is a recent technique for compiling quantum algorithms that simplifies transport operations by leveraging the concept of parity tracking~\cite{Klaver2024}. It is inspired by the Parity Architecture \cite{lechner2015quantum, fellner2022universal}, and optimized for quantum chips with low connectivity. The key concept of the Parity Architecture is to encode the relative magnetization of two logical qubits $i$ and $j$ (referred to as the \textit{parity} of the qubits) in a physical qubit $(i, j)$, termed \textit{parity qubit}. While this introduces an overhead in the number of qubits, it allows for the efficient implementation of some classes of quantum algorithm by employing single-qubit $Z$-rotations on parity qubits, which correspond to logical two-qubit operations~\cite{lechner2020quantum, fellner2022applications}. In the original Parity Architecture, the parity of all pairs of logical qubits $i$ and $j$ is encoded in a physical qubit, which ensures that the state of the parity qubits is accessible at any point during the execution of a quantum algorithm. Instead of using all parity qubits simultaneously, the Parity Flow approach allows to operate on a subset of parity qubits that are encoded on demand depending on the gates to be implemented. The Parity Flow formalism can be used to compile circuits for devices with linear nearest-neighbor (LNN) connectivity, a process referred to as Linear Parity Compilation (LPC). The primary advantage of LPC is that all operations required for a quantum circuit, including both quantum gates and encoding steps, can be mapped onto a linear chain of qubits without the need for SWAP operations. This is especially beneficial for linear QCCD architectures, where SWAP operations are among the most time-consuming transport operations~\cite{pino2021demonstration}.

In this work, we analyze the impact of algorithmic optimizations inspired by the Parity Flow formalism on the runtime of quantum algorithms on linear QCCDss, and we compare our results with the circuits generated with the well-known compilers provided by TKET~\cite{Sivarajah_2021tket} and Qiskit~\cite{Qiskit}, for two different use cases: the Quantum Approximate Optimization Algorithm (QAOA) for all-to-all connected Hamiltonians and the quantum Fourier transform (QFT). QCCDs devices typically have native M{\o}lmer-S{\o}rensen (MS) gates~\cite{sorensen2000entanglement} which implement $R_{zz}(\theta)$ interactions. In devices with arbitrary angle interactions, the LPC requires twice as many $R_{zz}$ gates but, in turn, no SWAP operations, leading to the same total number of number of two-qubit operations ($R_{zz}$ plus SWAP gates) as other approaches. Since MS gates are considerably faster than SWAP gates~\cite{pino2021demonstration}, we estimate that the Parity Flow approach halves the execution time for QAOA and QFT circuits, in comparison to circuits obtained with TKET and Qiskit. Because arbitrary-angle interactions can be technically challenging~\cite{gerster2022experimental}, it is preferable for some platforms to decompose interactions into fully entangling (${\theta = \tfrac{\pi}{2}}$) MS gates. In that case, our algorithm proposals require the same number of $R_{zz}(\tfrac{\pi}{2})$ gates as the standard approaches, however, no SWAPs of ions, resulting in a considerable reduction in the total number of two-qubit interactions and a further reduction of the runtime. 

\section{Background}

\subsection{Linear QCCDs} \label{sec:linear qccds}
\begin{figure*}
    \centering
    \includegraphics[width=\textwidth]{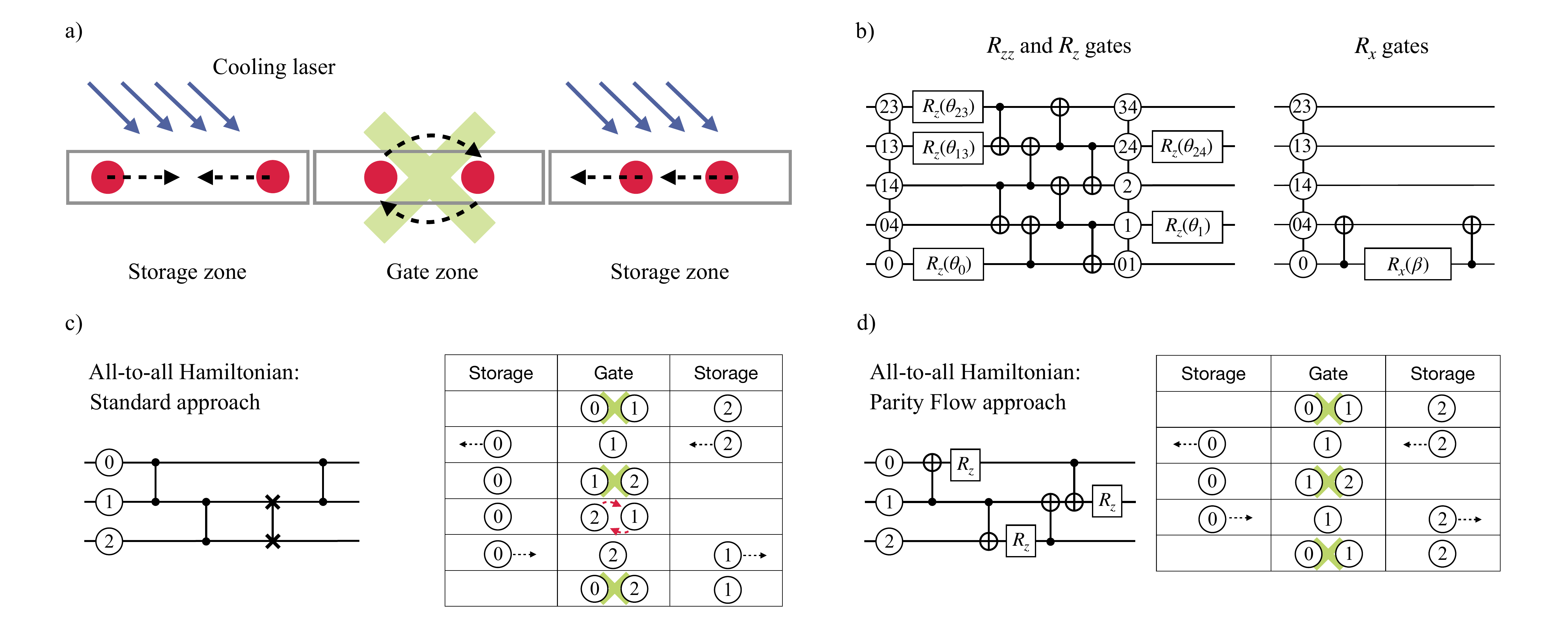}
    \caption{Swap-less algorithms for linear QCCDs using Parity Flow. 
    (a) QCCDs feature an interaction zone, where high-power laser light is available (green X), enabling the implementation of quantum gates, and several storage zones. To execute a quantum circuit, ions are transported and sorted by linear shifts (dashed straight arrows). Further available operations are merging/splitting independent ion chains, and swapping ions (dashed curved arrows). Ions are selectively moved to the gate zone, to which SWAP operations are usually restricted. Cooling lasers, available in all zones of the device, reduce the temperature of the ions, which increases due to transport operations. 
    (b) $R_{zz}$ gates are implemented as single-qubit rotations on the corresponding parity qubit $(i,j)$ (left panel). As an example, we show the logical gates $\tilde{R}_{z_2,z_3}$, $\tilde{R}_{z_2,z_4}$ and $\tilde{R}_{z_1,z_3}$. The necessary parity qubits are encoded on demand using CNOT gates [see Eqs.~\eqref{eq:encode} and~\eqref{eq:remove}]. Any parity qubit can be constructed using first-neighbor interactions in a linear chain of ions. Single-qubit rotations are executed on the target logical qubits $i$, like $\tilde{R}_{z_0}$ and $\tilde{R}_{z_1}$ in the figure. $R_x$ gates require at most two CNOT gates to collect the necessary logical information onto one physical qubit. In the right image, we show the gate sequence for an $\tilde{R}_{x_0}$ rotation. The vertical lines connecting the qubits correspond to spanning lines that contain all the information to reconstruct the logical state.
    (c) Transport operations to implement an all-to-all, Ising-type Hamiltonian in a three-qubit system. Swapping ions is fundamental to achieve all-to-all connectivity. 
    (d) Implementation of the same algorithm using the Parity Flow formalism. The total number of two-qubit operations is the same, but SWAP gates are eliminated, which reduces the algorithmic runtime (see Sec.~\ref{sec: runtime reduction}).}
    \label{fig:encoding_decoding}
\end{figure*}

Linear traps are a common approach in trapped-ion quantum computing~\cite{pino2021demonstration, moses2023race, Wright2019benchmarking, pogorelov2021compact, kaushal2020shuttling, piltz2016varsatile, Hahn2019integrated}. Qubits are encoded in ions, which are held in place by electromagnetic fields and manipulated using laser or microwave pulses. However, the number of trapped ions cannot increase indefinitely, because a large number of trapped ions induces a structural transition in the ion chain~\cite{fishman2008structural}. QCCDs~\cite{Kielpinski2002} circumvent this problem by employing segmented electrodes, which are individually controlled. In this way, ions are organized in several small ion crystals, that can be transported and resorted. The fundamental concepts of linear QCCDs are depicted in Fig.~\ref{fig:encoding_decoding}(a). The quantum chip is designed to have a trap zone were laser light is available (known as the interaction or the gate zone), and therefore gates can be implemented by moving the ions to this zone~\cite{kaushal2020shuttling, pino2021demonstration}. Furthermore, there are several zones for ion storage. 

The ability to transport ions between different regions of the trap is fundamental to achieve all-to-all connectivity between the qubits. Transport is enabled by adjusting the voltages on electrodes that form the trapping potential. The primitive transport operations are (i) linear shifts, to move ions along the trap, (ii) merges, to combine two ion crystals into a single one, (iii) splits, to separate an ion crystal into two smaller crystals and (iv) physical SWAPs, to interchange two ions in the chain~\cite{kaushal2020shuttling, pino2021demonstration, moses2023race}. 

Transport operations inevitably increase the ion chain temperature by exciting the transverse and longitudinal vibrational normal modes~\cite{pino2021demonstration}. Although the MS gate is temperature-resistant, increasing temperature deteriorates the gate fidelity~\cite{sorensen2000entanglement}. Therefore cooling operations (including Doppler- and side-band cooling~\cite{wayne1995cooling}) must be applied during quantum circuit execution. To protect the quantum state during cooling, the ion crystals are cooled using sympathetic cooling techniques~\cite{mao2021experimental, barrett2003sympathetic}. Because heating is partially induced by transport, the reduction of transport operations will also reduce the cooling requirements.

Combined transport and cooling operations dominate the runtime of a circuit, accounting for 98 to 99 percent of the total time~\cite{moses2023race}. QCCDs usually employ the hyperfine energy levels of an ion to encode the qubit levels, which can result in coherence times as long as $5500 \si{\second}$~\cite{Wang2021single}. However, achieving such long coherence times requires complex control techniques. Current reported values of the coherence time for commercial QCCDs are in the order of seconds~\cite{pino2021demonstration}, which is already the order of the runtime of 32-qubit circuits~\cite{moses2023race}. Therefore, scaling the number of qubits requires increasing the coherence time or simplifying the transport operations, to reduce the transport and cooling time of the circuit. 

Among the primitive transport operations, physical SWAPs are the most time-consuming and mainly responsible for the heating of the chain~\cite{pino2021demonstration}. SWAP gates require independent control of all involved trap electrodes~\cite{splatt2009deterministic, kaushal2020shuttling}, a requirement that cannot be satisfied on arbitrarily large chips due to the large number of control signals involved. Therefore, SWAP gates are restricted to specific zones in the chip in some platforms, causing additional transport costs~\cite{moses2023race, pino2021demonstration}.

\subsection{Linear Parity Compilation}\label{sec:linear_parity}

Linear Parity Compilation (LPC) uses the principles of Parity Flow tracking~\cite{Klaver2024} to compile arbitrary quantum circuits to a linear chain of qubits with linear nearest-neighbor connectivity (LNN). Given a quantum circuit of $n$ logical qubits, LPC requires the same number $n$ of physical qubits to encode the quantum state. In the computational basis, each logical qubit $(i)$ is represented by a superposition of the binary values ${s_i=0,1}$ that can be encoded using a single physical qubit. We call these qubits \textit{base} qubits. Physical qubits can also encode the parity of two arbitrary logical qubits ${(i,j)}$ as a superposition of the binary sum of the two eigenvalues, ${s_i \oplus s_j}$. LPC combines both base and parity qubits to represent the logical state at any time. If every logical qubit index $i$ appears at least once (in a parity qubit $(i,j)$ or directly encoded as a base qubit) and if at least one logical qubit is encoded as a base qubit, then the full logical state of all qubits can be deduced from the encoded qubits. Any set of $n$ qubits satisfying these conditions is called a \textit{spanning line}. 

The parity information can be encoded and decoded using CNOT gates, acting as
\begin{equation} \label{eq:encode}
    \mathrm{CNOT}: \; |s_i\rangle_c |s_j\rangle_t \rightarrow |s_i\rangle_c |s_i \oplus s_j\rangle_t
\end{equation}
in the computational basis, which describes the creation of parity qubit $(i,j)$. The subscripts $c$ and $t$ indicate control and target qubit, respectively. In the same way, the logical information can be recovered from the parity qubits:
\begin{equation} \label{eq:remove}
    \mathrm{CNOT}: \; |s_i\rangle_c |s_i \oplus s_j\rangle_t \rightarrow |s_i\rangle_c |s_j \rangle_t.
\end{equation}
Logical two-body $ZZ$ gates on an encoded state $|\psi_c\rangle$, are applied according to
\begin{equation}
    \label{eq:logical_z}
    \tilde{Z_{i}} \tilde{Z_{j}}|\psi_c\rangle=Z_{i,j}|\psi_c\rangle,
\end{equation}
where the physical $Z_{i,j}$ operator is localized to a single parity qubit. Here and in the following, logical operators are denoted with a tilde. 

An arbitrary quantum algorithm can be decomposed into an ordered sequence of logical quantum gates in the universal gate set $\{ \tilde{R}_{z_i,z_j}, \tilde{R}_{z_i}, \tilde{R}_{x_i} \}$. To operate the logical $Z$-basis gates $\tilde{R}_{z_i,z_j}$ and $\tilde{R}_{z_i}$, single-qubit $Z$-rotation are applied on the corresponding parity qubit $(i,j)$ or base qubit $(i)$, in accordance with Eq.~\eqref{eq:logical_z}. The idea of LPC essentially reduces to encoding one spanning line at a time. If the qubits required to implement certain gates are not present in the current spanning line, CNOT gates are used to transform a spanning line into another by using the CNOT sequence depicted in Fig.~\ref{fig:encoding_decoding}(b), until the desired qubit appears. Note that all possible parity qubits $(i,j)$ and base qubits $(i)$ will eventually appear. Therefore, single and two-qubit logical $Z$-basis operations can be implemented as a gate on a single physical qubit throughout the procedure.

To apply a logical $X$-basis gate, the parity information of the target qubit $(i)$ must be localized in a single qubit. The logical information can be removed from additional qubits using CNOT gates as in Eq.~\eqref{eq:remove}. Once the target qubit state is present in one single physical qubit [which can be a base qubit $(i)$ or a parity qubit $(i,j)$], the $X$-gate can be applied, see Fig.~\ref{fig:encoding_decoding}(b). After applying the $X$-rotation, the qubit information has to be restored by an additional CNOT gate. By construction of the spanning lines, quantum $X$-information about a single logical qubit is only spread over at most two neighboring physical qubits. Therefore, an $X$-gate can be applied with at most two CNOT gates at any point in the algorithm.

In Fig.~\ref{fig:encoding_decoding}(c) and (d) we exemplify the execution of an all-to-all interacting Hamiltonian ${H=\sum_{i,j}\theta_{ij}\sigma_z^{(i)}\sigma_z^{(j)}}$ using the standard approach (with SWAP gates) and Parity Flow, for a small circuit (${n=3}$). In both cases, the total number of two-qubit operations is the same, but Parity Flow approach does not require SWAP gates. This property is corroborated for larger systems in Sec.~\ref{sec:resources}.

\section{Compilation for a linear QCCD} \label{sec:resources}

The LPC is particularly beneficial for algorithms with Ising-type interactions, $\sum_{i,j}J_{i,j}\sigma_z^{(i)}\sigma_z^{(j)}$. In this section, we present the gate budget for a single round of QAOA (including the mixer and the phase separator layers) and for the QFT when implementing the approach using the LPC as proposed in Ref.~\cite{Klaver2024}. We compare the LPC results with the gate counts obtained with the Qiskit~\cite{Qiskit} and TKET~\cite{Sivarajah_2021tket} compilers.

\subsection{Quantum Approximate Optimization}

The QAOA was introduced by E. Farhi et al.~\cite{farhi2014quantum} and is considered a promising candidate to show quantum advantage for optimization problems in the NISQ-era~\cite{preskill2018quantum}, especially after many variations have been introduced~\cite{Hadfield2019, blekos2023review}. It represents a hybrid quantum-classical heuristic to solve combinatorial optimization problems, encoded in an Ising-like problem Hamiltonian
\begin{equation} \label{eq:qaoa_ham}
    H_\text{P}=\sum_{j=1}^n\sum_{i<j} J_{ij}\sigma_z^{(i)}\sigma_z^{(j)}+\sum_{i=1}^n h_i \sigma_z^{(i)}
\end{equation}
on $n$ qubits, where $J_{ij}$ are coupling strengths between qubits and $h_i$ denote local fields. The QAOA prepares a candidate state 
\begin{equation}
    \label{eq:qaoa_state}
    \ket{\psi(\bm{\beta}, \bm{\gamma})} = \prod_{j=1}^p e^{-i\beta_j H_\text{X}} e^{-i\gamma_j H_\text{P}}\ket{+}^{\otimes n},
\end{equation}
where ${H_X=\sum_{i=1}^n \sigma_x^{(i)}}$ is a transversal driving term. This candidate state is then optimized by varying the $2p$ parameters ${\bm{\beta}=(\beta_1, \dots, \beta_p)}$ and ${\bm{\gamma}=(\gamma_1, \dots, \gamma_p)}$ in a quantum-classical feedback loop, seeking $\min_{\bm{\beta},\bm{\gamma}}\braket{\psi(\bm{\beta}, \bm{\gamma})|H_\text{P}|\psi(\bm{\beta}, \bm{\gamma})}$.

To implement all terms in $H_\text{P}$, the circuit obtained via LPC sequentially transforms an initial spanning line into ${n-1}$ other spanning lines resulting in a CNOT depth of ${2(n-1)}$ and a CNOT gate count of ${(n-1)^{2}}$~\cite{Klaver2024}, which is sufficient to access all possible two-party parity qubits and base qubits. For every spanning line, the parity- and base qubits can be rotated according to $R_{z}(J_{ij})$ and $R_{z}(h_i)$ in order to prepare the candidate state in Eq.~\eqref{eq:qaoa_state}. 

The operations corresponding to $H_{X}$ can be applied using logical $R_{x}(\beta_{j})$ gates as shown in Fig.~\ref{fig:encoding_decoding}(b). By construction of the spanning lines, the logical $X$-information of any logical qubit at a given time always spans over at most two neighboring qubits and a single logical $X$-rotation can be implemented with at most two CNOT gates, as discussed in Sec.~\ref{sec:linear_parity}. Implementing this rotation for all logical qubits requires a constant circuit depth of $4$ and at most ${2(n-2)}$ CNOT gates. In total, we therefore obtain a gate count of ${n^2-1}$~\cite{Klaver2024}. CNOT gates can be implemented using a single $R_{zz}$ gate of angle $\tfrac{\pi}{2} $ and additional single-qubit gates. Therefore the two-qubit gate count does not change when expressing the circuit in terms of $R_{zz}$ gates. 

\subsection{Quantum Fourier Transform}
The QFT is a subroutine for many corner-stone quantum algorithms such as Shor's factoring algorithm~\cite{shor1999polynomial}. It consists of a series of Hadamard ($\mathrm{H}$) and $\mathrm{CPhase}$ gates, as shown in Algorithm~\ref{alg:qft}.

\begin{algorithm}[H]
\caption{Quantum Fourier Transform (QFT)}
\label{alg:qft}
\begin{algorithmic}[1]
\State \textbf{Input:} $n$-qubit quantum state $\ket{\psi} = \ket{\psi_0, \psi_1, \dots, \psi_{n-1}}$
\State \textbf{Output:} Quantum state transformed by QFT
\Procedure{QFT}{$n$}
    \For{$i = 0$ to $n-1$}
        \State Apply $\mathrm{H}(i)$ 
        \For{$j = i+1$ to $n-1$}
            \State Apply $\mathrm{CPhase}(\frac{\pi}{2^{j-i}})$ between qubits $i$ and $j$
        \EndFor
    \EndFor
    \State Reverse the order of the qubits
\EndProcedure
\end{algorithmic}
\end{algorithm}

Using a parity-based algorithm, it is possible to implement the QFT on an LNN device consuming ${n^2-1}$ CNOT gates and a total circuit depth of ${5n-3}$ (${4n-4}$ taking only two-qubit gates into account)~\cite{Klaver2024}. The scheme starts with a set of qubits holding the initial quantum state and then applies the QFT in the Parity Architecture as presented in Ref.~\cite{fellner2022applications}, adapted to the linear chain~\cite{Klaver2024}. The CPhase gates in the QFT circuit are decomposed into three $R_z$ rotations on data- and parity qubits according to 
\begin{equation}
     \text{CP}_{ij}(\phi) = R_{z_{i}}\left(\frac{\phi}{2}\right)R_{z_{i}z_{j}}\left(-\frac{\phi}{2}\right)R_{z_{j}}\left(\frac{\phi}{2}\right),
\end{equation}
as derived in Ref.~\cite{fellner2022universal}. The Hadamard gates on qubits ${2,\dots, n-1}$ are decomposed into $R_z$ and $R_x$ gates, whereas the ones on qubits $1$ and $n$ can be applied directly, because there are no interfering CPhase gates before and after the gate, respectively. The parity-based algorithm intrinsically reverses the order of the qubits such that line 10 in Algorithm~\ref{alg:qft} can be omitted, which saves ${\lceil n/2\rceil}$ if all-to-all connectivity is provided or $\mathcal{O}(n^2)$ SWAP gates on an LNN device.

\subsection{Comparison with standard compilers} \label{sect:comparison_standard_compiler}

We calculate the gate resources for the execution of the algorithms on a linear QCCD, using the Qiskit~\cite{Qiskit} and TKET~\cite{Sivarajah_2021tket} compilers. We consider a linear device with LNN connectivity. We compile to SWAP gates (which are applied by means of physical exchange of ions), arbitrary-angle $R_{zz}(\theta)$ gates and single-qubit gates. For the compilation with Qiskit (version 1.2.0) we use the maximum optimization level and average over $500$ realizations. In the case of TKET, which yields a deterministic output, we use version 1.22.0, and heavy-optimization passes such as \texttt{FullPeepholeOptimise} and \texttt{CliffordSimp}~\cite{Sivarajah_2021tket}. 

\begin{figure*}
    \centering
    \includegraphics[width=1.\textwidth]{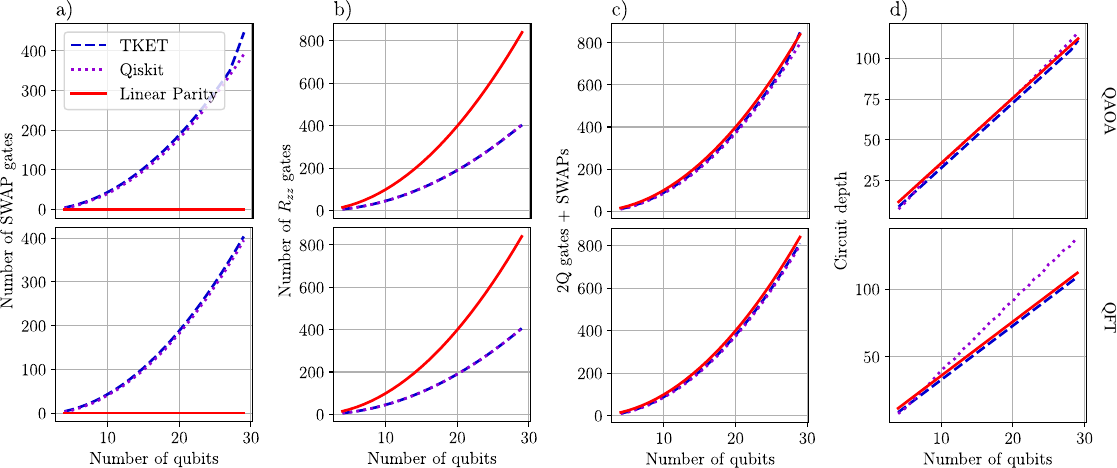}
    \caption{Resource estimation to run the QAOA (first row of panels) and the QFT (second row of panels) as a function of the number of qubits. We assume a quantum device with LNN connectivity, native $R_{zz}(\theta)$ gates and physical SWAPs. (a) Both the Qiskit and TKET compiler require $~\tfrac{n^2}{2}$ SWAPs, while the Parity-based algorithm does not require SWAPs. (b) LPC circuits require twice as many $R_{zz}$ gates as standard circuits, but (c) the total number of two-qubit operations to be performed in the gate zones ($R_{zz}$ and SWAPs) is the same in leading order for the three methods. (d) The circuit depth considering only two-qubit gates. For all approaches, the circuit depth increases linearly with $n$ but the slope is larger for QFT using Qiskit. These values may increase depending on the number of gate zones of the QCCD. In devices were only fully entangling gates $R_{zz}(\tfrac{\pi}{2})$ are available, the number of $R_{zz}$ gates for TKET and Qiskit doubles. In that case, LPC offers a significant reduction of the circuit depth and the number of two-qubit operations.} 
    \label{fig:resources}
\end{figure*}

The comparison between the LPC and the compilation with Qiskit and TKET is summarized in Fig.~\ref{fig:resources} for QAOA and QFT. For both algorithms, the major advantage of the LPC is the elimination of the SWAP gates, which correspond to $\tfrac{n(n-1)}{2} - 2$ elementary two-qubit operations, where $n$ is the number of logical qubits. The number of $R_{zz}(\theta)$ gates for the circuits compiled with standard approaches is $\tfrac{n(n-1)}{2}$, which is roughly half of the ${n^2-1}$ gates required for LPC-QAOA or LPC-QFT. The circuit depth for QFT is approximately the same using any of the three compilation strategies. Note that the fact that LPC-QAOA eliminates the need for rearranging the qubits at the end of the QFT (line 10 in Algorithm~\ref{alg:qft}) was not taken into account in our analysis. In the case of QAOA, TKET and LPC also yield similar results, but Qiskit performs worse for large $n$.

If only fully entangling gates $R_{zz}(\tfrac{\pi}{2})$ are available, then $R_{zz}(\theta)$ gates are decomposed into two $R_{zz}(\tfrac{\pi}{2})$ and additional single-qubit gates. This doubles the gate count for Qiskit and TKET but does not affect the circuits obtained via LPC, which are already expressed in terms of $R_{zz}(\tfrac{\pi}{2})$ gates. In this case, the total number of two-qubit operations and the circuit depth are considerably smaller for LPC, with the additional advantage of eliminating the SWAP gates. Although MS gates can in principle be implemented for any angle, the calibration of a MS gate is expensive, as it involves to determine several parameters that are non-linearly correlated~\cite{gerster2022experimental}. Arbitrary-angle MS gates are therefore expensive to calibrate and may not be available on all ion-based quantum computers.

\section{Runtime reduction for SWAP-less circuits} \label{sec: runtime reduction}

We finally estimate the reduction in the runtime when using the SWAP-less LPC. In linear QCCD platforms, physical SWAPs of ions are essential to allow all-to-all connectivity of qubits. 
As explained in Sec.~\ref{sec:linear qccds}, SWAP gates are among the most time-consuming transport operations in linear QCCDs. A SWAP-less circuit reduces the runtime by eliminating the time of the SWAP gates, but also by reducing the need of cooling and the shuttle. We estimate the order of magnitude of the transport and cooling time for an LPC circuit compared to the time required for standard QAOA and QFT compilations, to show that eliminating SWAP gates leads to a non-negligible reduction in the execution time of a circuit.  

The total runtime of a circuit depends on the details of the QCCD architecture. In this work, we focus on Quantinuum's H2 hardware~\cite{moses2023race}, a QCCD platform with a closed-loop shape capable of holding up to 50 qubits \cite{decross2024computational}. For this hardware, the runtime is reasonably well predicted with the formula used to determine the cost of a circuit run, defined in terms of H-System Quantum Credits ($HQC$)~\cite{racetrack2024sheet} as
\begin{equation} \label{eq: quantinuum_hqc}
    HQC = 5 + C \frac{n_{1q} + 10n_{2q} + 5 M}{5000},
\end{equation}
where $C$ is the number of shots, $n_{1,2q}$ are the number of single- and two-qubit gates and $M$ the number of state-preparation and measurement operations. For the circuits we consider here, ${10n_{2q} \gg n_{1q} + 5 M}$ and therefore ${HQC \sim 0.002 n_{2q}}$ for a single shot holds. An $HQC$ takes typically between 3 and 7 seconds to execute, so the runtime is
\begin{equation} \label{eq:runtime_h2}
    T_{\mathrm{run}} (n_{2q}) \approx 0.01 n_{2q} 	\si{\second},
\end{equation}
which we obtain by assuming an average runtime of 5 seconds for one HQC.

\begin{figure}
    \centering
    \includegraphics[width=1.\columnwidth]{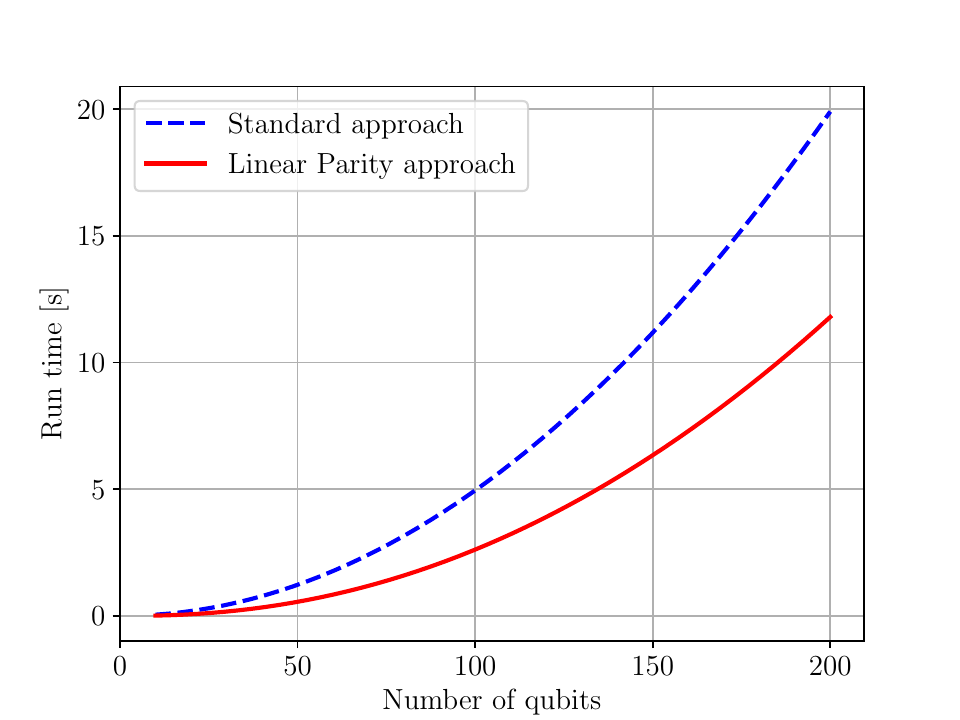}
    \caption{Runtime comparison for algorithms obtained via LPC and via standard compilers, considering the gate counts from Fig.~\ref{fig:resources} and assuming a time cost of $1\si{ms}$ per SWAP operation. The runtime for algorithms based on LPC is around half of the runtime using a standard compiler. The total runtime includes both transport and cooling times. The transport time is estimated from the number of two-qubit gates using the HQC formula for Quantinuum's H-Systems [see Eq.~\eqref{eq: quantinuum_hqc}] while the cooling time is estimated using Eq.~\eqref{eq:cooling} based on the circuit time budget from Ref.~\cite{moses2023race}. The scaling law for the cooling time is derived from experimental data, for circuits with 16 and 32 qubits, and with transport times within the same values as those analyzed here (see Appendix \ref{app:cooling}).}
    \label{fig:run_time}
\end{figure}

The runtime is dominated by the cooling and the transport time $T_{\mathrm{run}} (n_{2q}) = T_{\mathrm{transp}} + T_{\mathrm{cool}}$. The cooling time is estimated from the transport time (see Appendix~\ref{app:cooling}),
\begin{equation} \label{eq:cooling}
    T_{\mathrm{cool}} = 0.6\; \left(\tfrac{T_{\mathrm{transp}}}{\si{\second}}\right)^{0.9} 	\si{\second},
\end{equation}
from which we determine the transport time $T_{\mathrm{transp}}$ as a function of $n_{2q}$ by the equation 
\begin{equation} \label{eq:transport_time}
    T_{\mathrm{transp}} + 0.6\; \left(\tfrac{T_{\mathrm{transp}}}{\si{\second}}\right)^{0.9} \si{\second}= 0.01 n_{2q} 	\si{\second}. 
\end{equation}
We now estimate the reduction of $ T_{\mathrm{run}}$ for algorithms compiled with LPC. Equation~\eqref{eq:transport_time} gives the transport time per two-qubit gate including all possible transport operations, such as linear shifts, merge and split operations, in ion chains and physical SWAP gates. For LPC, we assume a reduction in the transport time proportional to the number of SWAP gates $n_{\mathrm{SWAP}}$:
\begin{equation} \label{eq:transport_lp}
    T_{\mathrm{transp}}^{(LP)} = T_{\mathrm{transp}} - \alpha \;n_{\mathrm{SWAP}},
\end{equation}
where $\alpha$ is the average time cost per SWAP. We use the results from Section \ref{sect:comparison_standard_compiler} for the gate counting. We obtain $n_{2q} =\tfrac{n(n-1)}{2}$ for standard compilers (Qiskit and TKET), and twice this value for LPC and additional $n_{\mathrm{SWAP}}=\tfrac{n^2}{2} - \tfrac{n}{2} $ SWAP gates for the standard compilers. The time cost $\alpha$ per SWAP operation can be estimated from the time budget in Table II of Ref.~\cite{pino2021demonstration}. A physical SWAP of ions takes $200 \si{\us}$, but additional transport time is required, as ions must be in a gate zone to perform the SWAP, and ion chains must be combined into a single chain. Because a shift can take around $300	\si{\us}$ and the combine-and-split operations take also around $300\si{\us}$, we estimate the time cost per SWAP $\alpha$ to be in the order of $1\si{\ms}$. In Fig.~\ref{fig:run_time}, we show the runtime as a function of the number of qubits $n$, for circuits with ${n_{2q} = \tfrac{n^2}{2} - \tfrac{n}{2} - 2}$ (like QAOA and QFT), using Eq.~\eqref{eq:runtime_h2}. We also plot the estimated runtime for LPC algorithms, using the reduced transport time $T_{\mathrm{transp}}^{(LP)}$ from Eq.~\eqref{eq:transport_lp}. For the current capacity of Quantinuum's H2 (50 qubits), the estimated runtime is ${1.2 \si{\second}}$ using a standard compilation, while the estimated runtime for LPC algorithms is ${0.5 \si{\second}}$, due to the reduction in transport and cooling time. 

\section{Conclusions}
Our findings demonstrate that the Parity Flow approach for linear architectures significantly reduces the algorithmic runtime for QFT and QAOA on linear QCCDs. This reduction is important for scaling, as it allows for an increase in the number of quantum gates that can be operated within the coherence time. The algorithms currently implemented on QCCDs operate at runtimes that already push the boundaries of what coherence time allows. Therefore, scaling up the number of qubits necessitates extending coherence time, a complex engineering challenge. The development of optimized algorithms like LPC that reduce the need for transport operations allow for a greater number of qubits within the current coherence time constraints. Our estimations indicate that circuits using LPC will be approximately twice as fast compared as the circuits obtained via the Qiskit and TKET compilers, which allows the number of qubits to be increased by a factor of $\sqrt{2}$, since the runtime scales approximately quadratic with the number of qubits $n$. With only 200 qubits, QCCDs projected to surpass classical computers in certain QAOA use cases~\cite{ruslan2024evidence}. A single QAOA round for an all-to-all connected problem on 200 qubits would require at least 20 seconds of coherence with Qiskit and TKET compilers, while only 12 seconds are enough when using the optimized approach presented in Ref.~\cite{Klaver2024}.

An additional advantage of LPC algorithms is that the interactions are problem-independent, meaning that two-qubit gates only need to be optimized for the fully entangling angle of $\tfrac{\pi}{2}$. Compared to other fixed-angle decompositions, LPC algorithms offer a significant reduction in the number of two-qubit operations (including $R_{zz}$ and SWAP gates). Since multi-qubit gate calibration is a complex and time-consuming task~\cite{gerster2022experimental} that must be performed for every interaction zone, fixed-angle algorithms simplify this process by requiring calibration for only a single angle.

With the development of appropriate algorithms, linear QCCDs have the potential to evolve from proof-of-principle hardware into economically viable and competitive alternatives to classical algorithms. The local connectivity of LPC algorithms simplifies the necessary transport and thus shortens the circuit execution time, allowing deeper circuits to be executed with the same coherence time. This advancement could position QCCDs as not only relevant in quantum research but also practical for a broad range of applications within the quantum computing industry.

\section*{Acknowledgements}

The authors thank Dr.~Adu Offei-Danso for fruitful discussions. 
This study was supported by the Austrian Research Promotion Agency (FFG Project No. FO999909249, FFG Basisprogramm).
This research was funded in part by the Austrian Science Fund (FWF) under Grant-DOI 10.55776/F71 and Grant-DOI 10.55776/Y1067.
This project was funded within the QuantERA II Programme that has received funding from the European Union’s Horizon 2020 research and innovation programme under Grant Agreement No. 101017733. This publication has received funding under Horizon Europe Programme HORIZON-CL4-2022-QUANTUM-02-SGA via the project 101113690 (PASQuanS2.1)
We acknowledge funding within the ATIQ project supported by the Federal Ministry of Education and Research (BMBF) within the framework of the program ``Quantum technologies - from basic research to market'' under Grant No. 13N16127. 
For the purpose of open access, the author has applied a CC BY public copyright license to any Author Accepted Manuscript version arising from this submission.

\appendix

\section{Cooling time} \label{app:cooling}
Doppler and sideband cooling are mid-circuit operations in QCCDs, necessary for maintaining high fidelity of quantum gates. Because transport operations increase the kinetic energy (and therefore the temperature) of ion chains, the cooling time increases with the transport requirements of a circuit. In this Appendix, we estimate the relationship between cooling time and transport time. 

\begin{figure}
    \centering
    \includegraphics[width=1\columnwidth]{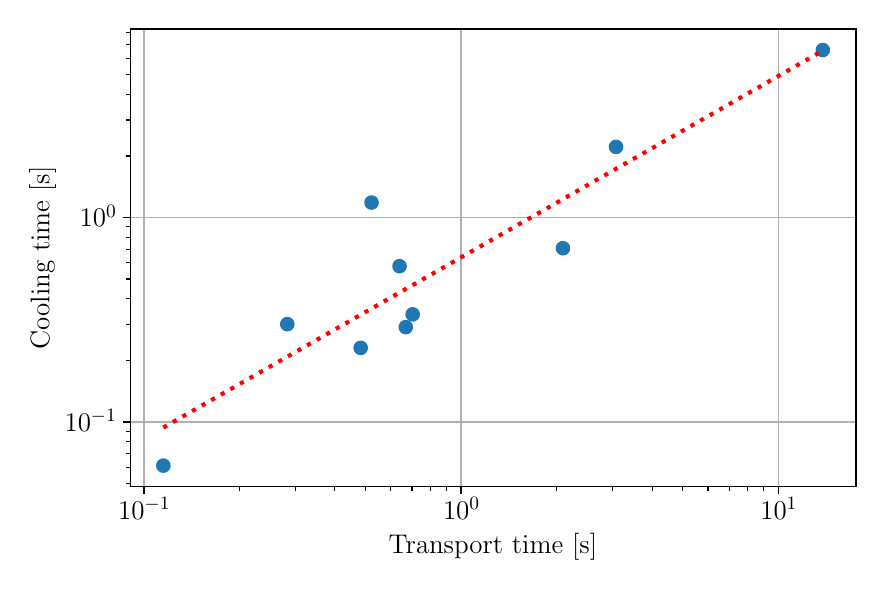}
    \caption{Cooling time as a function of the transport time in a linear QCCD. Blue dots are experimental data from Ref.~\cite{moses2023race} and the dotted red line is a power function fit $y = a\left(\tfrac{x}{\si{\second}}\right)^b$, with ${a = (0.6 \pm 0.1)\si{\second}}$  and ${b=(0.9 \pm 0.1)}$. We use this relation to interpolate the cooling time from the transport time.}
    \label{fig:fit_fig}
\end{figure}

We base our analysis on the circuit budget presented in Table I of Ref.~\cite{moses2023race}, which provides transport, cooling, and gate times for various example circuits. In Fig.~\ref{fig:fit_fig}, we plot the cooling time as a function of transport time for these circuits and identify a power-law relationship as described in Eq.~\eqref{eq:cooling}. The circuits considered in this plot are for algorithms with 32 and 16 qubits. We assume that this power-law relationship holds for any qubit system size, based on the following argument. Transport operations can be applied in parallel, meaning the same transport time can introduce more energy into the ion system as the number of ions increases. However, since cooling can also be performed simultaneously on multiple qubits, we assume the relationship between cooling and transport times remains valid across different system sizes.

It is important to note that the scaling law presented in Eq.~\eqref{eq:cooling} is not a general formula for calculating the cooling time. Instead, it serves as an approximate expression to interpolate cooling time within the analyzed range of transport times, specifically between $0.1$ and $10 \si{s}$, which is the range considered in Fig.~\ref{fig:run_time}.

\bibliography{bibliography}

\begin{thebibliography}{38}%
\makeatletter
\providecommand \@ifxundefined [1]{%
 \@ifx{#1\undefined}
}%
\providecommand \@ifnum [1]{%
 \ifnum #1\expandafter \@firstoftwo
 \else \expandafter \@secondoftwo
 \fi
}%
\providecommand \@ifx [1]{%
 \ifx #1\expandafter \@firstoftwo
 \else \expandafter \@secondoftwo
 \fi
}%
\providecommand \natexlab [1]{#1}%
\providecommand \enquote  [1]{``#1''}%
\providecommand \bibnamefont  [1]{#1}%
\providecommand \bibfnamefont [1]{#1}%
\providecommand \citenamefont [1]{#1}%
\providecommand \href@noop [0]{\@secondoftwo}%
\providecommand \href [0]{\begingroup \@sanitize@url \@href}%
\providecommand \@href[1]{\@@startlink{#1}\@@href}%
\providecommand \@@href[1]{\endgroup#1\@@endlink}%
\providecommand \@sanitize@url [0]{\catcode `\\12\catcode `\$12\catcode `\&12\catcode `\#12\catcode `\^12\catcode `\_12\catcode `\%12\relax}%
\providecommand \@@startlink[1]{}%
\providecommand \@@endlink[0]{}%
\providecommand \url  [0]{\begingroup\@sanitize@url \@url }%
\providecommand \@url [1]{\endgroup\@href {#1}{\urlprefix }}%
\providecommand \urlprefix  [0]{URL }%
\providecommand \Eprint [0]{\href }%
\providecommand \doibase [0]{http://dx.doi.org/}%
\providecommand \selectlanguage [0]{\@gobble}%
\providecommand \bibinfo  [0]{\@secondoftwo}%
\providecommand \bibfield  [0]{\@secondoftwo}%
\providecommand \translation [1]{[#1]}%
\providecommand \BibitemOpen [0]{}%
\providecommand \bibitemStop [0]{}%
\providecommand \bibitemNoStop [0]{.\EOS\space}%
\providecommand \EOS [0]{\spacefactor3000\relax}%
\providecommand \BibitemShut  [1]{\csname bibitem#1\endcsname}%
\let\auto@bib@innerbib\@empty
\bibitem [{\citenamefont {Klaver}\ \emph {et~al.}(2024)\citenamefont {Klaver}, \citenamefont {Rombouts}, \citenamefont {Fellner}, \citenamefont {Messinger}, \citenamefont {Ender}, \citenamefont {Ludwig},\ and\ \citenamefont {Lechner}}]{Klaver2024}%
  \BibitemOpen
  \bibfield  {author} {\bibinfo {author} {\bibfnamefont {Berend}\ \bibnamefont {Klaver}}, \bibinfo {author} {\bibfnamefont {Stefan}\ \bibnamefont {Rombouts}}, \bibinfo {author} {\bibfnamefont {Michael}\ \bibnamefont {Fellner}}, \bibinfo {author} {\bibfnamefont {Anette}\ \bibnamefont {Messinger}}, \bibinfo {author} {\bibfnamefont {Kilian}\ \bibnamefont {Ender}}, \bibinfo {author} {\bibfnamefont {Katharina}\ \bibnamefont {Ludwig}}, \ and\ \bibinfo {author} {\bibfnamefont {Wolfgang}\ \bibnamefont {Lechner}},\ }\href@noop {} {\enquote {\bibinfo {title} {Swap-less implementation of quantum algorithms},}\ } (\bibinfo {year} {2024}),\ \Eprint {http://arxiv.org/abs/2408.10907} {arXiv:2408.10907 [quant-ph]} \BibitemShut {NoStop}%
\bibitem [{\citenamefont {Kielpinski}\ \emph {et~al.}(2002)\citenamefont {Kielpinski}, \citenamefont {Monroe},\ and\ \citenamefont {Wineland}}]{Kielpinski2002}%
  \BibitemOpen
  \bibfield  {author} {\bibinfo {author} {\bibfnamefont {D.}~\bibnamefont {Kielpinski}}, \bibinfo {author} {\bibfnamefont {C.}~\bibnamefont {Monroe}}, \ and\ \bibinfo {author} {\bibfnamefont {D.~J.}\ \bibnamefont {Wineland}},\ }\bibfield  {title} {\enquote {\bibinfo {title} {Architecture for a large-scale ion-trap quantum computer},}\ }\href {\doibase 10.1038/nature00784} {\bibfield  {journal} {\bibinfo  {journal} {Nature}\ }\textbf {\bibinfo {volume} {417}},\ \bibinfo {pages} {709--711} (\bibinfo {year} {2002})}\BibitemShut {NoStop}%
\bibitem [{\citenamefont {Moses}\ \emph {et~al.}(2023)\citenamefont {Moses}, \citenamefont {Baldwin}, \citenamefont {Allman}, \citenamefont {Ancona}, \citenamefont {Ascarrunz}, \citenamefont {Barnes}, \citenamefont {Bartolotta}, \citenamefont {Bjork}, \citenamefont {Blanchard}, \citenamefont {Bohn}, \citenamefont {Bohnet}, \citenamefont {Brown}, \citenamefont {Burdick}, \citenamefont {Burton}, \citenamefont {Campbell}, \citenamefont {Campora}, \citenamefont {Carron}, \citenamefont {Chambers}, \citenamefont {Chan}, \citenamefont {Chen}, \citenamefont {Chernoguzov}, \citenamefont {Chertkov}, \citenamefont {Colina}, \citenamefont {Curtis}, \citenamefont {Daniel}, \citenamefont {DeCross}, \citenamefont {Deen}, \citenamefont {Delaney}, \citenamefont {Dreiling}, \citenamefont {Ertsgaard}, \citenamefont {Esposito}, \citenamefont {Estey}, \citenamefont {Fabrikant}, \citenamefont {Figgatt}, \citenamefont {Foltz}, \citenamefont {Foss-Feig}, \citenamefont {Francois}, \citenamefont {Gaebler}, \citenamefont {Gatterman},
  \citenamefont {Gilbreth}, \citenamefont {Giles}, \citenamefont {Glynn}, \citenamefont {Hall}, \citenamefont {Hankin}, \citenamefont {Hansen}, \citenamefont {Hayes}, \citenamefont {Higashi}, \citenamefont {Hoffman}, \citenamefont {Horning}, \citenamefont {Hout}, \citenamefont {Jacobs}, \citenamefont {Johansen}, \citenamefont {Jones}, \citenamefont {Karcz}, \citenamefont {Klein}, \citenamefont {Lauria}, \citenamefont {Lee}, \citenamefont {Liefer}, \citenamefont {Lu}, \citenamefont {Lucchetti}, \citenamefont {Lytle}, \citenamefont {Malm}, \citenamefont {Matheny}, \citenamefont {Mathewson}, \citenamefont {Mayer}, \citenamefont {Miller}, \citenamefont {Mills}, \citenamefont {Neyenhuis}, \citenamefont {Nugent}, \citenamefont {Olson}, \citenamefont {Parks}, \citenamefont {Price}, \citenamefont {Price}, \citenamefont {Pugh}, \citenamefont {Ransford}, \citenamefont {Reed}, \citenamefont {Roman}, \citenamefont {Rowe}, \citenamefont {Ryan-Anderson}, \citenamefont {Sanders}, \citenamefont {Sedlacek}, \citenamefont
  {Shevchuk}, \citenamefont {Siegfried}, \citenamefont {Skripka}, \citenamefont {Spaun}, \citenamefont {Sprenkle}, \citenamefont {Stutz}, \citenamefont {Swallows}, \citenamefont {Tobey}, \citenamefont {Tran}, \citenamefont {Tran}, \citenamefont {Vogt}, \citenamefont {Volin}, \citenamefont {Walker}, \citenamefont {Zolot},\ and\ \citenamefont {Pino}}]{moses2023race}%
  \BibitemOpen
  \bibfield  {author} {\bibinfo {author} {\bibfnamefont {S.~A.}\ \bibnamefont {Moses}}, \bibinfo {author} {\bibfnamefont {C.~H.}\ \bibnamefont {Baldwin}}, \bibinfo {author} {\bibfnamefont {M.~S.}\ \bibnamefont {Allman}}, \bibinfo {author} {\bibfnamefont {R.}~\bibnamefont {Ancona}}, \bibinfo {author} {\bibfnamefont {L.}~\bibnamefont {Ascarrunz}}, \bibinfo {author} {\bibfnamefont {C.}~\bibnamefont {Barnes}}, \bibinfo {author} {\bibfnamefont {J.}~\bibnamefont {Bartolotta}}, \bibinfo {author} {\bibfnamefont {B.}~\bibnamefont {Bjork}}, \bibinfo {author} {\bibfnamefont {P.}~\bibnamefont {Blanchard}}, \bibinfo {author} {\bibfnamefont {M.}~\bibnamefont {Bohn}}, \bibinfo {author} {\bibfnamefont {J.~G.}\ \bibnamefont {Bohnet}}, \bibinfo {author} {\bibfnamefont {N.~C.}\ \bibnamefont {Brown}}, \bibinfo {author} {\bibfnamefont {N.~Q.}\ \bibnamefont {Burdick}}, \bibinfo {author} {\bibfnamefont {W.~C.}\ \bibnamefont {Burton}}, \bibinfo {author} {\bibfnamefont {S.~L.}\ \bibnamefont {Campbell}}, \bibinfo {author}
  {\bibfnamefont {J.~P.}\ \bibnamefont {Campora}}, \bibinfo {author} {\bibfnamefont {C.}~\bibnamefont {Carron}}, \bibinfo {author} {\bibfnamefont {J.}~\bibnamefont {Chambers}}, \bibinfo {author} {\bibfnamefont {J.~W.}\ \bibnamefont {Chan}}, \bibinfo {author} {\bibfnamefont {Y.~H.}\ \bibnamefont {Chen}}, \bibinfo {author} {\bibfnamefont {A.}~\bibnamefont {Chernoguzov}}, \bibinfo {author} {\bibfnamefont {E.}~\bibnamefont {Chertkov}}, \bibinfo {author} {\bibfnamefont {J.}~\bibnamefont {Colina}}, \bibinfo {author} {\bibfnamefont {J.~P.}\ \bibnamefont {Curtis}}, \bibinfo {author} {\bibfnamefont {R.}~\bibnamefont {Daniel}}, \bibinfo {author} {\bibfnamefont {M.}~\bibnamefont {DeCross}}, \bibinfo {author} {\bibfnamefont {D.}~\bibnamefont {Deen}}, \bibinfo {author} {\bibfnamefont {C.}~\bibnamefont {Delaney}}, \bibinfo {author} {\bibfnamefont {J.~M.}\ \bibnamefont {Dreiling}}, \bibinfo {author} {\bibfnamefont {C.~T.}\ \bibnamefont {Ertsgaard}}, \bibinfo {author} {\bibfnamefont {J.}~\bibnamefont {Esposito}}, \bibinfo
  {author} {\bibfnamefont {B.}~\bibnamefont {Estey}}, \bibinfo {author} {\bibfnamefont {M.}~\bibnamefont {Fabrikant}}, \bibinfo {author} {\bibfnamefont {C.}~\bibnamefont {Figgatt}}, \bibinfo {author} {\bibfnamefont {C.}~\bibnamefont {Foltz}}, \bibinfo {author} {\bibfnamefont {M.}~\bibnamefont {Foss-Feig}}, \bibinfo {author} {\bibfnamefont {D.}~\bibnamefont {Francois}}, \bibinfo {author} {\bibfnamefont {J.~P.}\ \bibnamefont {Gaebler}}, \bibinfo {author} {\bibfnamefont {T.~M.}\ \bibnamefont {Gatterman}}, \bibinfo {author} {\bibfnamefont {C.~N.}\ \bibnamefont {Gilbreth}}, \bibinfo {author} {\bibfnamefont {J.}~\bibnamefont {Giles}}, \bibinfo {author} {\bibfnamefont {E.}~\bibnamefont {Glynn}}, \bibinfo {author} {\bibfnamefont {A.}~\bibnamefont {Hall}}, \bibinfo {author} {\bibfnamefont {A.~M.}\ \bibnamefont {Hankin}}, \bibinfo {author} {\bibfnamefont {A.}~\bibnamefont {Hansen}}, \bibinfo {author} {\bibfnamefont {D.}~\bibnamefont {Hayes}}, \bibinfo {author} {\bibfnamefont {B.}~\bibnamefont {Higashi}}, \bibinfo
  {author} {\bibfnamefont {I.~M.}\ \bibnamefont {Hoffman}}, \bibinfo {author} {\bibfnamefont {B.}~\bibnamefont {Horning}}, \bibinfo {author} {\bibfnamefont {J.~J.}\ \bibnamefont {Hout}}, \bibinfo {author} {\bibfnamefont {R.}~\bibnamefont {Jacobs}}, \bibinfo {author} {\bibfnamefont {J.}~\bibnamefont {Johansen}}, \bibinfo {author} {\bibfnamefont {L.}~\bibnamefont {Jones}}, \bibinfo {author} {\bibfnamefont {J.}~\bibnamefont {Karcz}}, \bibinfo {author} {\bibfnamefont {T.}~\bibnamefont {Klein}}, \bibinfo {author} {\bibfnamefont {P.}~\bibnamefont {Lauria}}, \bibinfo {author} {\bibfnamefont {P.}~\bibnamefont {Lee}}, \bibinfo {author} {\bibfnamefont {D.}~\bibnamefont {Liefer}}, \bibinfo {author} {\bibfnamefont {S.~T.}\ \bibnamefont {Lu}}, \bibinfo {author} {\bibfnamefont {D.}~\bibnamefont {Lucchetti}}, \bibinfo {author} {\bibfnamefont {C.}~\bibnamefont {Lytle}}, \bibinfo {author} {\bibfnamefont {A.}~\bibnamefont {Malm}}, \bibinfo {author} {\bibfnamefont {M.}~\bibnamefont {Matheny}}, \bibinfo {author} {\bibfnamefont
  {B.}~\bibnamefont {Mathewson}}, \bibinfo {author} {\bibfnamefont {K.}~\bibnamefont {Mayer}}, \bibinfo {author} {\bibfnamefont {D.~B.}\ \bibnamefont {Miller}}, \bibinfo {author} {\bibfnamefont {M.}~\bibnamefont {Mills}}, \bibinfo {author} {\bibfnamefont {B.}~\bibnamefont {Neyenhuis}}, \bibinfo {author} {\bibfnamefont {L.}~\bibnamefont {Nugent}}, \bibinfo {author} {\bibfnamefont {S.}~\bibnamefont {Olson}}, \bibinfo {author} {\bibfnamefont {J.}~\bibnamefont {Parks}}, \bibinfo {author} {\bibfnamefont {G.~N.}\ \bibnamefont {Price}}, \bibinfo {author} {\bibfnamefont {Z.}~\bibnamefont {Price}}, \bibinfo {author} {\bibfnamefont {M.}~\bibnamefont {Pugh}}, \bibinfo {author} {\bibfnamefont {A.}~\bibnamefont {Ransford}}, \bibinfo {author} {\bibfnamefont {A.~P.}\ \bibnamefont {Reed}}, \bibinfo {author} {\bibfnamefont {C.}~\bibnamefont {Roman}}, \bibinfo {author} {\bibfnamefont {M.}~\bibnamefont {Rowe}}, \bibinfo {author} {\bibfnamefont {C.}~\bibnamefont {Ryan-Anderson}}, \bibinfo {author} {\bibfnamefont
  {S.}~\bibnamefont {Sanders}}, \bibinfo {author} {\bibfnamefont {J.}~\bibnamefont {Sedlacek}}, \bibinfo {author} {\bibfnamefont {P.}~\bibnamefont {Shevchuk}}, \bibinfo {author} {\bibfnamefont {P.}~\bibnamefont {Siegfried}}, \bibinfo {author} {\bibfnamefont {T.}~\bibnamefont {Skripka}}, \bibinfo {author} {\bibfnamefont {B.}~\bibnamefont {Spaun}}, \bibinfo {author} {\bibfnamefont {R.~T.}\ \bibnamefont {Sprenkle}}, \bibinfo {author} {\bibfnamefont {R.~P.}\ \bibnamefont {Stutz}}, \bibinfo {author} {\bibfnamefont {M.}~\bibnamefont {Swallows}}, \bibinfo {author} {\bibfnamefont {R.~I.}\ \bibnamefont {Tobey}}, \bibinfo {author} {\bibfnamefont {A.}~\bibnamefont {Tran}}, \bibinfo {author} {\bibfnamefont {T.}~\bibnamefont {Tran}}, \bibinfo {author} {\bibfnamefont {E.}~\bibnamefont {Vogt}}, \bibinfo {author} {\bibfnamefont {C.}~\bibnamefont {Volin}}, \bibinfo {author} {\bibfnamefont {J.}~\bibnamefont {Walker}}, \bibinfo {author} {\bibfnamefont {A.~M.}\ \bibnamefont {Zolot}}, \ and\ \bibinfo {author} {\bibfnamefont
  {J.~M.}\ \bibnamefont {Pino}},\ }\bibfield  {title} {\enquote {\bibinfo {title} {A race-track trapped-ion quantum processor},}\ }\href {\doibase 10.1103/PhysRevX.13.041052} {\bibfield  {journal} {\bibinfo  {journal} {Phys. Rev. X}\ }\textbf {\bibinfo {volume} {13}},\ \bibinfo {pages} {041052} (\bibinfo {year} {2023})}\BibitemShut {NoStop}%
\bibitem [{\citenamefont {Lekitsch}\ \emph {et~al.}(2017)\citenamefont {Lekitsch}, \citenamefont {Weidt}, \citenamefont {Fowler}, \citenamefont {Mølmer}, \citenamefont {Devitt}, \citenamefont {Wunderlich},\ and\ \citenamefont {Hensinger}}]{Bjoern2017blueprint}%
  \BibitemOpen
  \bibfield  {author} {\bibinfo {author} {\bibfnamefont {Bjoern}\ \bibnamefont {Lekitsch}}, \bibinfo {author} {\bibfnamefont {Sebastian}\ \bibnamefont {Weidt}}, \bibinfo {author} {\bibfnamefont {Austin~G.}\ \bibnamefont {Fowler}}, \bibinfo {author} {\bibfnamefont {Klaus}\ \bibnamefont {Mølmer}}, \bibinfo {author} {\bibfnamefont {Simon~J.}\ \bibnamefont {Devitt}}, \bibinfo {author} {\bibfnamefont {Christof}\ \bibnamefont {Wunderlich}}, \ and\ \bibinfo {author} {\bibfnamefont {Winfried~K.}\ \bibnamefont {Hensinger}},\ }\bibfield  {title} {\enquote {\bibinfo {title} {Blueprint for a microwave trapped ion quantum computer},}\ }\href {\doibase 10.1126/sciadv.1601540} {\bibfield  {journal} {\bibinfo  {journal} {Science Advances}\ }\textbf {\bibinfo {volume} {3}},\ \bibinfo {pages} {e1601540} (\bibinfo {year} {2017})}\BibitemShut {NoStop}%
\bibitem [{\citenamefont {Malinowski}\ \emph {et~al.}(2023)\citenamefont {Malinowski}, \citenamefont {Allcock},\ and\ \citenamefont {Ballance}}]{malinowski2023howtowire}%
  \BibitemOpen
  \bibfield  {author} {\bibinfo {author} {\bibfnamefont {M.}~\bibnamefont {Malinowski}}, \bibinfo {author} {\bibfnamefont {D.T.C.}\ \bibnamefont {Allcock}}, \ and\ \bibinfo {author} {\bibfnamefont {C.J.}\ \bibnamefont {Ballance}},\ }\bibfield  {title} {\enquote {\bibinfo {title} {How to wire a $1000$-qubit trapped-ion quantum computer},}\ }\href {\doibase 10.1103/PRXQuantum.4.040313} {\bibfield  {journal} {\bibinfo  {journal} {PRX Quantum}\ }\textbf {\bibinfo {volume} {4}},\ \bibinfo {pages} {040313} (\bibinfo {year} {2023})}\BibitemShut {NoStop}%
\bibitem [{\citenamefont {Weidt}\ \emph {et~al.}(2016)\citenamefont {Weidt}, \citenamefont {Randall}, \citenamefont {Webster}, \citenamefont {Lake}, \citenamefont {Webb}, \citenamefont {Cohen}, \citenamefont {Navickas}, \citenamefont {Lekitsch}, \citenamefont {Retzker},\ and\ \citenamefont {Hensinger}}]{weidt2016trapped}%
  \BibitemOpen
  \bibfield  {author} {\bibinfo {author} {\bibfnamefont {S.}~\bibnamefont {Weidt}}, \bibinfo {author} {\bibfnamefont {J.}~\bibnamefont {Randall}}, \bibinfo {author} {\bibfnamefont {S.~C.}\ \bibnamefont {Webster}}, \bibinfo {author} {\bibfnamefont {K.}~\bibnamefont {Lake}}, \bibinfo {author} {\bibfnamefont {A.~E.}\ \bibnamefont {Webb}}, \bibinfo {author} {\bibfnamefont {I.}~\bibnamefont {Cohen}}, \bibinfo {author} {\bibfnamefont {T.}~\bibnamefont {Navickas}}, \bibinfo {author} {\bibfnamefont {B.}~\bibnamefont {Lekitsch}}, \bibinfo {author} {\bibfnamefont {A.}~\bibnamefont {Retzker}}, \ and\ \bibinfo {author} {\bibfnamefont {W.~K.}\ \bibnamefont {Hensinger}},\ }\bibfield  {title} {\enquote {\bibinfo {title} {Trapped-ion quantum logic with global radiation fields},}\ }\href {\doibase 10.1103/PhysRevLett.117.220501} {\bibfield  {journal} {\bibinfo  {journal} {Phys. Rev. Lett.}\ }\textbf {\bibinfo {volume} {117}},\ \bibinfo {pages} {220501} (\bibinfo {year} {2016})}\BibitemShut {NoStop}%
\bibitem [{\citenamefont {Kaushal}\ \emph {et~al.}(2020)\citenamefont {Kaushal}, \citenamefont {Lekitsch}, \citenamefont {Stahl}, \citenamefont {Hilder}, \citenamefont {Pijn}, \citenamefont {Schmiegelow}, \citenamefont {Bermudez}, \citenamefont {M{\"u}ller}, \citenamefont {Schmidt-Kaler},\ and\ \citenamefont {Poschinger}}]{kaushal2020shuttling}%
  \BibitemOpen
  \bibfield  {author} {\bibinfo {author} {\bibfnamefont {Vidyut}\ \bibnamefont {Kaushal}}, \bibinfo {author} {\bibfnamefont {Bjoern}\ \bibnamefont {Lekitsch}}, \bibinfo {author} {\bibfnamefont {A}~\bibnamefont {Stahl}}, \bibinfo {author} {\bibfnamefont {J}~\bibnamefont {Hilder}}, \bibinfo {author} {\bibfnamefont {Daniel}\ \bibnamefont {Pijn}}, \bibinfo {author} {\bibfnamefont {C}~\bibnamefont {Schmiegelow}}, \bibinfo {author} {\bibfnamefont {Alejandro}\ \bibnamefont {Bermudez}}, \bibinfo {author} {\bibfnamefont {M}~\bibnamefont {M{\"u}ller}}, \bibinfo {author} {\bibfnamefont {Ferdinand}\ \bibnamefont {Schmidt-Kaler}}, \ and\ \bibinfo {author} {\bibfnamefont {U}~\bibnamefont {Poschinger}},\ }\bibfield  {title} {\enquote {\bibinfo {title} {Shuttling-based trapped-ion quantum information processing},}\ }\href {\doibase https://doi.org/10.1116/1.5126186} {\bibfield  {journal} {\bibinfo  {journal} {AVS Quantum Science}\ }\textbf {\bibinfo {volume} {2}},\ \bibinfo {pages} {014101} (\bibinfo {year}
  {2020})}\BibitemShut {NoStop}%
\bibitem [{\citenamefont {Shaydulin}\ \emph {et~al.}(2024)\citenamefont {Shaydulin}, \citenamefont {Li}, \citenamefont {Chakrabarti}, \citenamefont {DeCross}, \citenamefont {Herman}, \citenamefont {Kumar}, \citenamefont {Larson}, \citenamefont {Lykov}, \citenamefont {Minssen}, \citenamefont {Sun}, \citenamefont {Alexeev}, \citenamefont {Dreiling}, \citenamefont {Gaebler}, \citenamefont {Gatterman}, \citenamefont {Gerber}, \citenamefont {Gilmore}, \citenamefont {Gresh}, \citenamefont {Hewitt}, \citenamefont {Horst}, \citenamefont {Hu}, \citenamefont {Johansen}, \citenamefont {Matheny}, \citenamefont {Mengle}, \citenamefont {Mills}, \citenamefont {Moses}, \citenamefont {Neyenhuis}, \citenamefont {Siegfried}, \citenamefont {Yalovetzky},\ and\ \citenamefont {Pistoia}}]{ruslan2024evidence}%
  \BibitemOpen
  \bibfield  {author} {\bibinfo {author} {\bibfnamefont {Ruslan}\ \bibnamefont {Shaydulin}}, \bibinfo {author} {\bibfnamefont {Changhao}\ \bibnamefont {Li}}, \bibinfo {author} {\bibfnamefont {Shouvanik}\ \bibnamefont {Chakrabarti}}, \bibinfo {author} {\bibfnamefont {Matthew}\ \bibnamefont {DeCross}}, \bibinfo {author} {\bibfnamefont {Dylan}\ \bibnamefont {Herman}}, \bibinfo {author} {\bibfnamefont {Niraj}\ \bibnamefont {Kumar}}, \bibinfo {author} {\bibfnamefont {Jeffrey}\ \bibnamefont {Larson}}, \bibinfo {author} {\bibfnamefont {Danylo}\ \bibnamefont {Lykov}}, \bibinfo {author} {\bibfnamefont {Pierre}\ \bibnamefont {Minssen}}, \bibinfo {author} {\bibfnamefont {Yue}\ \bibnamefont {Sun}}, \bibinfo {author} {\bibfnamefont {Yuri}\ \bibnamefont {Alexeev}}, \bibinfo {author} {\bibfnamefont {Joan~M.}\ \bibnamefont {Dreiling}}, \bibinfo {author} {\bibfnamefont {John~P.}\ \bibnamefont {Gaebler}}, \bibinfo {author} {\bibfnamefont {Thomas~M.}\ \bibnamefont {Gatterman}}, \bibinfo {author} {\bibfnamefont {Justin~A.}\
  \bibnamefont {Gerber}}, \bibinfo {author} {\bibfnamefont {Kevin}\ \bibnamefont {Gilmore}}, \bibinfo {author} {\bibfnamefont {Dan}\ \bibnamefont {Gresh}}, \bibinfo {author} {\bibfnamefont {Nathan}\ \bibnamefont {Hewitt}}, \bibinfo {author} {\bibfnamefont {Chandler~V.}\ \bibnamefont {Horst}}, \bibinfo {author} {\bibfnamefont {Shaohan}\ \bibnamefont {Hu}}, \bibinfo {author} {\bibfnamefont {Jacob}\ \bibnamefont {Johansen}}, \bibinfo {author} {\bibfnamefont {Mitchell}\ \bibnamefont {Matheny}}, \bibinfo {author} {\bibfnamefont {Tanner}\ \bibnamefont {Mengle}}, \bibinfo {author} {\bibfnamefont {Michael}\ \bibnamefont {Mills}}, \bibinfo {author} {\bibfnamefont {Steven~A.}\ \bibnamefont {Moses}}, \bibinfo {author} {\bibfnamefont {Brian}\ \bibnamefont {Neyenhuis}}, \bibinfo {author} {\bibfnamefont {Peter}\ \bibnamefont {Siegfried}}, \bibinfo {author} {\bibfnamefont {Romina}\ \bibnamefont {Yalovetzky}}, \ and\ \bibinfo {author} {\bibfnamefont {Marco}\ \bibnamefont {Pistoia}},\ }\bibfield  {title} {\enquote {\bibinfo
  {title} {Evidence of scaling advantage for the quantum approximate optimization algorithm on a classically intractable problem},}\ }\href {\doibase 10.1126/sciadv.adm6761} {\bibfield  {journal} {\bibinfo  {journal} {Science Advances}\ }\textbf {\bibinfo {volume} {10}},\ \bibinfo {pages} {eadm6761} (\bibinfo {year} {2024})}\BibitemShut {NoStop}%
\bibitem [{\citenamefont {Monroe}\ and\ \citenamefont {Kim}(2013)}]{monroe2013scaling}%
  \BibitemOpen
  \bibfield  {author} {\bibinfo {author} {\bibfnamefont {C.}~\bibnamefont {Monroe}}\ and\ \bibinfo {author} {\bibfnamefont {J.}~\bibnamefont {Kim}},\ }\bibfield  {title} {\enquote {\bibinfo {title} {Scaling the ion trap quantum processor},}\ }\href {\doibase 10.1126/science.1231298} {\bibfield  {journal} {\bibinfo  {journal} {Science}\ }\textbf {\bibinfo {volume} {339}},\ \bibinfo {pages} {1164--1169} (\bibinfo {year} {2013})}\BibitemShut {NoStop}%
\bibitem [{\citenamefont {Pino}\ \emph {et~al.}(2021)\citenamefont {Pino}, \citenamefont {Dreiling}, \citenamefont {Figgatt}, \citenamefont {Gaebler}, \citenamefont {Moses}, \citenamefont {Allman}, \citenamefont {Baldwin}, \citenamefont {Foss-Feig}, \citenamefont {Hayes}, \citenamefont {Mayer}, \citenamefont {Ryan-Anderson},\ and\ \citenamefont {Neyenhuis}}]{pino2021demonstration}%
  \BibitemOpen
  \bibfield  {author} {\bibinfo {author} {\bibfnamefont {J.~M.}\ \bibnamefont {Pino}}, \bibinfo {author} {\bibfnamefont {J.~M.}\ \bibnamefont {Dreiling}}, \bibinfo {author} {\bibfnamefont {C.}~\bibnamefont {Figgatt}}, \bibinfo {author} {\bibfnamefont {J.~P.}\ \bibnamefont {Gaebler}}, \bibinfo {author} {\bibfnamefont {S.~A.}\ \bibnamefont {Moses}}, \bibinfo {author} {\bibfnamefont {M.~S.}\ \bibnamefont {Allman}}, \bibinfo {author} {\bibfnamefont {C.~H.}\ \bibnamefont {Baldwin}}, \bibinfo {author} {\bibfnamefont {M.}~\bibnamefont {Foss-Feig}}, \bibinfo {author} {\bibfnamefont {D.}~\bibnamefont {Hayes}}, \bibinfo {author} {\bibfnamefont {K.}~\bibnamefont {Mayer}}, \bibinfo {author} {\bibfnamefont {C.}~\bibnamefont {Ryan-Anderson}}, \ and\ \bibinfo {author} {\bibfnamefont {B.}~\bibnamefont {Neyenhuis}},\ }\bibfield  {title} {\enquote {\bibinfo {title} {Demonstration of the trapped-ion quantum ccd computer architecture},}\ }\href {\doibase 10.1038/s41586-021-03318-4} {\bibfield  {journal} {\bibinfo  {journal}
  {Nature}\ }\textbf {\bibinfo {volume} {592}},\ \bibinfo {pages} {209--213} (\bibinfo {year} {2021})}\BibitemShut {NoStop}%
\bibitem [{\citenamefont {Palani}\ \emph {et~al.}(2023)\citenamefont {Palani}, \citenamefont {Hasse}, \citenamefont {Kiefer}, \citenamefont {Boeckling}, \citenamefont {Schroeder}, \citenamefont {Warring},\ and\ \citenamefont {Schaetz}}]{palani2023high}%
  \BibitemOpen
  \bibfield  {author} {\bibinfo {author} {\bibfnamefont {Deviprasath}\ \bibnamefont {Palani}}, \bibinfo {author} {\bibfnamefont {Florian}\ \bibnamefont {Hasse}}, \bibinfo {author} {\bibfnamefont {Philip}\ \bibnamefont {Kiefer}}, \bibinfo {author} {\bibfnamefont {Frederick}\ \bibnamefont {Boeckling}}, \bibinfo {author} {\bibfnamefont {Jan-Philipp}\ \bibnamefont {Schroeder}}, \bibinfo {author} {\bibfnamefont {Ulrich}\ \bibnamefont {Warring}}, \ and\ \bibinfo {author} {\bibfnamefont {Tobias}\ \bibnamefont {Schaetz}},\ }\bibfield  {title} {\enquote {\bibinfo {title} {High-fidelity transport of trapped-ion qubits in a multilayer array},}\ }\href {\doibase 10.1103/PhysRevA.107.L050601} {\bibfield  {journal} {\bibinfo  {journal} {Phys. Rev. A}\ }\textbf {\bibinfo {volume} {107}},\ \bibinfo {pages} {L050601} (\bibinfo {year} {2023})}\BibitemShut {NoStop}%
\bibitem [{\citenamefont {Kaufmann}\ \emph {et~al.}(2018)\citenamefont {Kaufmann}, \citenamefont {Gloger}, \citenamefont {Kaufmann}, \citenamefont {Johanning},\ and\ \citenamefont {Wunderlich}}]{kaufmann2018high}%
  \BibitemOpen
  \bibfield  {author} {\bibinfo {author} {\bibfnamefont {Peter}\ \bibnamefont {Kaufmann}}, \bibinfo {author} {\bibfnamefont {Timm~F.}\ \bibnamefont {Gloger}}, \bibinfo {author} {\bibfnamefont {Delia}\ \bibnamefont {Kaufmann}}, \bibinfo {author} {\bibfnamefont {Michael}\ \bibnamefont {Johanning}}, \ and\ \bibinfo {author} {\bibfnamefont {Christof}\ \bibnamefont {Wunderlich}},\ }\bibfield  {title} {\enquote {\bibinfo {title} {High-fidelity preservation of quantum information during trapped-ion transport},}\ }\href {\doibase 10.1103/PhysRevLett.120.010501} {\bibfield  {journal} {\bibinfo  {journal} {Phys. Rev. Lett.}\ }\textbf {\bibinfo {volume} {120}},\ \bibinfo {pages} {010501} (\bibinfo {year} {2018})}\BibitemShut {NoStop}%
\bibitem [{\citenamefont {Blakestad}\ \emph {et~al.}(2009)\citenamefont {Blakestad}, \citenamefont {Ospelkaus}, \citenamefont {VanDevender}, \citenamefont {Amini}, \citenamefont {Britton}, \citenamefont {Leibfried},\ and\ \citenamefont {Wineland}}]{Blakestad2009high}%
  \BibitemOpen
  \bibfield  {author} {\bibinfo {author} {\bibfnamefont {R.~B.}\ \bibnamefont {Blakestad}}, \bibinfo {author} {\bibfnamefont {C.}~\bibnamefont {Ospelkaus}}, \bibinfo {author} {\bibfnamefont {A.~P.}\ \bibnamefont {VanDevender}}, \bibinfo {author} {\bibfnamefont {J.~M.}\ \bibnamefont {Amini}}, \bibinfo {author} {\bibfnamefont {J.}~\bibnamefont {Britton}}, \bibinfo {author} {\bibfnamefont {D.}~\bibnamefont {Leibfried}}, \ and\ \bibinfo {author} {\bibfnamefont {D.~J.}\ \bibnamefont {Wineland}},\ }\bibfield  {title} {\enquote {\bibinfo {title} {High-fidelity transport of trapped-ion qubits through an $\mathbf{X}$-junction trap array},}\ }\href {\doibase 10.1103/PhysRevLett.102.153002} {\bibfield  {journal} {\bibinfo  {journal} {Phys. Rev. Lett.}\ }\textbf {\bibinfo {volume} {102}},\ \bibinfo {pages} {153002} (\bibinfo {year} {2009})}\BibitemShut {NoStop}%
\bibitem [{\citenamefont {Lechner}\ \emph {et~al.}(2015)\citenamefont {Lechner}, \citenamefont {Hauke},\ and\ \citenamefont {Zoller}}]{lechner2015quantum}%
  \BibitemOpen
  \bibfield  {author} {\bibinfo {author} {\bibfnamefont {Wolfgang}\ \bibnamefont {Lechner}}, \bibinfo {author} {\bibfnamefont {Philipp}\ \bibnamefont {Hauke}}, \ and\ \bibinfo {author} {\bibfnamefont {Peter}\ \bibnamefont {Zoller}},\ }\bibfield  {title} {\enquote {\bibinfo {title} {A quantum annealing architecture with all-to-all connectivity from local interactions},}\ }\href {\doibase 10.1126/sciadv.1500838} {\bibfield  {journal} {\bibinfo  {journal} {Science Advances}\ }\textbf {\bibinfo {volume} {1}},\ \bibinfo {pages} {e1500838} (\bibinfo {year} {2015})}\BibitemShut {NoStop}%
\bibitem [{\citenamefont {Fellner}\ \emph {et~al.}(2022{\natexlab{a}})\citenamefont {Fellner}, \citenamefont {Messinger}, \citenamefont {Ender},\ and\ \citenamefont {Lechner}}]{fellner2022universal}%
  \BibitemOpen
  \bibfield  {author} {\bibinfo {author} {\bibfnamefont {Michael}\ \bibnamefont {Fellner}}, \bibinfo {author} {\bibfnamefont {Anette}\ \bibnamefont {Messinger}}, \bibinfo {author} {\bibfnamefont {Kilian}\ \bibnamefont {Ender}}, \ and\ \bibinfo {author} {\bibfnamefont {Wolfgang}\ \bibnamefont {Lechner}},\ }\bibfield  {title} {\enquote {\bibinfo {title} {Universal parity quantum computing},}\ }\href {\doibase 10.1103/PhysRevLett.129.180503} {\bibfield  {journal} {\bibinfo  {journal} {Phys. Rev. Lett.}\ }\textbf {\bibinfo {volume} {129}},\ \bibinfo {pages} {180503} (\bibinfo {year} {2022}{\natexlab{a}})}\BibitemShut {NoStop}%
\bibitem [{\citenamefont {Lechner}(2020)}]{lechner2020quantum}%
  \BibitemOpen
  \bibfield  {author} {\bibinfo {author} {\bibfnamefont {W}~\bibnamefont {Lechner}},\ }\bibfield  {title} {\enquote {\bibinfo {title} {Quantum {A}pproximate {O}ptimization {W}ith {P}arallelizable {G}ates},}\ }\href {\doibase 10.1109/TQE.2020.3034798} {\bibfield  {journal} {\bibinfo  {journal} {IEEE Trans. Quantum Eng.}\ }\textbf {\bibinfo {volume} {1}},\ \bibinfo {pages} {1--6} (\bibinfo {year} {2020})}\BibitemShut {NoStop}%
\bibitem [{\citenamefont {Fellner}\ \emph {et~al.}(2022{\natexlab{b}})\citenamefont {Fellner}, \citenamefont {Messinger}, \citenamefont {Ender},\ and\ \citenamefont {Lechner}}]{fellner2022applications}%
  \BibitemOpen
  \bibfield  {author} {\bibinfo {author} {\bibfnamefont {Michael}\ \bibnamefont {Fellner}}, \bibinfo {author} {\bibfnamefont {Anette}\ \bibnamefont {Messinger}}, \bibinfo {author} {\bibfnamefont {Kilian}\ \bibnamefont {Ender}}, \ and\ \bibinfo {author} {\bibfnamefont {Wolfgang}\ \bibnamefont {Lechner}},\ }\bibfield  {title} {\enquote {\bibinfo {title} {Applications of universal parity quantum computation},}\ }\href {\doibase 10.1103/PhysRevA.106.042442} {\bibfield  {journal} {\bibinfo  {journal} {Phys. Rev. A}\ }\textbf {\bibinfo {volume} {106}},\ \bibinfo {pages} {042442} (\bibinfo {year} {2022}{\natexlab{b}})}\BibitemShut {NoStop}%
\bibitem [{\citenamefont {Sivarajah}\ \emph {et~al.}(2020)\citenamefont {Sivarajah}, \citenamefont {Dilkes}, \citenamefont {Cowtan}, \citenamefont {Simmons}, \citenamefont {Edgington},\ and\ \citenamefont {Duncan}}]{Sivarajah_2021tket}%
  \BibitemOpen
  \bibfield  {author} {\bibinfo {author} {\bibfnamefont {Seyon}\ \bibnamefont {Sivarajah}}, \bibinfo {author} {\bibfnamefont {Silas}\ \bibnamefont {Dilkes}}, \bibinfo {author} {\bibfnamefont {Alexander}\ \bibnamefont {Cowtan}}, \bibinfo {author} {\bibfnamefont {Will}\ \bibnamefont {Simmons}}, \bibinfo {author} {\bibfnamefont {Alec}\ \bibnamefont {Edgington}}, \ and\ \bibinfo {author} {\bibfnamefont {Ross}\ \bibnamefont {Duncan}},\ }\bibfield  {title} {\enquote {\bibinfo {title} {t$|$ket⟩: a retargetable compiler for nisq devices},}\ }\href {\doibase 10.1088/2058-9565/ab8e92} {\bibfield  {journal} {\bibinfo  {journal} {Quantum Science and Technology}\ }\textbf {\bibinfo {volume} {6}},\ \bibinfo {pages} {014003} (\bibinfo {year} {2020})}\BibitemShut {NoStop}%
\bibitem [{\citenamefont {Javadi-Abhari}\ \emph {et~al.}(2024)\citenamefont {Javadi-Abhari}, \citenamefont {Treinish}, \citenamefont {Krsulich}, \citenamefont {Wood}, \citenamefont {Lishman}, \citenamefont {Gacon}, \citenamefont {Martiel}, \citenamefont {Nation}, \citenamefont {Bishop}, \citenamefont {Cross}, \citenamefont {Johnson},\ and\ \citenamefont {Gambetta}}]{Qiskit}%
  \BibitemOpen
  \bibfield  {author} {\bibinfo {author} {\bibfnamefont {Ali}\ \bibnamefont {Javadi-Abhari}}, \bibinfo {author} {\bibfnamefont {Matthew}\ \bibnamefont {Treinish}}, \bibinfo {author} {\bibfnamefont {Kevin}\ \bibnamefont {Krsulich}}, \bibinfo {author} {\bibfnamefont {Christopher~J.}\ \bibnamefont {Wood}}, \bibinfo {author} {\bibfnamefont {Jake}\ \bibnamefont {Lishman}}, \bibinfo {author} {\bibfnamefont {Julien}\ \bibnamefont {Gacon}}, \bibinfo {author} {\bibfnamefont {Simon}\ \bibnamefont {Martiel}}, \bibinfo {author} {\bibfnamefont {Paul~D.}\ \bibnamefont {Nation}}, \bibinfo {author} {\bibfnamefont {Lev~S.}\ \bibnamefont {Bishop}}, \bibinfo {author} {\bibfnamefont {Andrew~W.}\ \bibnamefont {Cross}}, \bibinfo {author} {\bibfnamefont {Blake~R.}\ \bibnamefont {Johnson}}, \ and\ \bibinfo {author} {\bibfnamefont {Jay~M.}\ \bibnamefont {Gambetta}},\ }\href@noop {} {\enquote {\bibinfo {title} {Quantum computing with qiskit},}\ } (\bibinfo {year} {2024}),\ \Eprint {http://arxiv.org/abs/2405.08810} {arXiv:2405.08810
  [quant-ph]} \BibitemShut {NoStop}%
\bibitem [{\citenamefont {S\o{}rensen}\ and\ \citenamefont {M\o{}lmer}(2000)}]{sorensen2000entanglement}%
  \BibitemOpen
  \bibfield  {author} {\bibinfo {author} {\bibfnamefont {Anders}\ \bibnamefont {S\o{}rensen}}\ and\ \bibinfo {author} {\bibfnamefont {Klaus}\ \bibnamefont {M\o{}lmer}},\ }\bibfield  {title} {\enquote {\bibinfo {title} {Entanglement and quantum computation with ions in thermal motion},}\ }\href {\doibase 10.1103/PhysRevA.62.022311} {\bibfield  {journal} {\bibinfo  {journal} {Phys. Rev. A}\ }\textbf {\bibinfo {volume} {62}},\ \bibinfo {pages} {022311} (\bibinfo {year} {2000})}\BibitemShut {NoStop}%
\bibitem [{\citenamefont {Gerster}\ \emph {et~al.}(2022)\citenamefont {Gerster}, \citenamefont {Mart\'{\i}nez-Garc\'{\i}a}, \citenamefont {Hrmo}, \citenamefont {van Mourik}, \citenamefont {Wilhelm}, \citenamefont {Vodola}, \citenamefont {M\"uller}, \citenamefont {Blatt}, \citenamefont {Schindler},\ and\ \citenamefont {Monz}}]{gerster2022experimental}%
  \BibitemOpen
  \bibfield  {author} {\bibinfo {author} {\bibfnamefont {Lukas}\ \bibnamefont {Gerster}}, \bibinfo {author} {\bibfnamefont {Fernando}\ \bibnamefont {Mart\'{\i}nez-Garc\'{\i}a}}, \bibinfo {author} {\bibfnamefont {Pavel}\ \bibnamefont {Hrmo}}, \bibinfo {author} {\bibfnamefont {Martin~W.}\ \bibnamefont {van Mourik}}, \bibinfo {author} {\bibfnamefont {Benjamin}\ \bibnamefont {Wilhelm}}, \bibinfo {author} {\bibfnamefont {Davide}\ \bibnamefont {Vodola}}, \bibinfo {author} {\bibfnamefont {Markus}\ \bibnamefont {M\"uller}}, \bibinfo {author} {\bibfnamefont {Rainer}\ \bibnamefont {Blatt}}, \bibinfo {author} {\bibfnamefont {Philipp}\ \bibnamefont {Schindler}}, \ and\ \bibinfo {author} {\bibfnamefont {Thomas}\ \bibnamefont {Monz}},\ }\bibfield  {title} {\enquote {\bibinfo {title} {Experimental bayesian calibration of trapped-ion entangling operations},}\ }\href {\doibase 10.1103/PRXQuantum.3.020350} {\bibfield  {journal} {\bibinfo  {journal} {PRX Quantum}\ }\textbf {\bibinfo {volume} {3}},\ \bibinfo {pages} {020350}
  (\bibinfo {year} {2022})}\BibitemShut {NoStop}%
\bibitem [{\citenamefont {Wright}\ \emph {et~al.}(2019)\citenamefont {Wright}, \citenamefont {Beck}, \citenamefont {Debnath}, \citenamefont {Amini}, \citenamefont {Nam}, \citenamefont {Grzesiak}, \citenamefont {Chen}, \citenamefont {Pisenti}, \citenamefont {Chmielewski}, \citenamefont {Collins}, \citenamefont {Hudek}, \citenamefont {Mizrahi}, \citenamefont {Wong-Campos}, \citenamefont {Allen}, \citenamefont {Apisdorf}, \citenamefont {Solomon}, \citenamefont {Williams}, \citenamefont {Ducore}, \citenamefont {Blinov}, \citenamefont {Kreikemeier}, \citenamefont {Chaplin}, \citenamefont {Keesan}, \citenamefont {Monroe},\ and\ \citenamefont {Kim}}]{Wright2019benchmarking}%
  \BibitemOpen
  \bibfield  {author} {\bibinfo {author} {\bibfnamefont {K.}~\bibnamefont {Wright}}, \bibinfo {author} {\bibfnamefont {K.~M.}\ \bibnamefont {Beck}}, \bibinfo {author} {\bibfnamefont {S.}~\bibnamefont {Debnath}}, \bibinfo {author} {\bibfnamefont {J.~M.}\ \bibnamefont {Amini}}, \bibinfo {author} {\bibfnamefont {Y.}~\bibnamefont {Nam}}, \bibinfo {author} {\bibfnamefont {N.}~\bibnamefont {Grzesiak}}, \bibinfo {author} {\bibfnamefont {J.-S.}\ \bibnamefont {Chen}}, \bibinfo {author} {\bibfnamefont {N.~C.}\ \bibnamefont {Pisenti}}, \bibinfo {author} {\bibfnamefont {M.}~\bibnamefont {Chmielewski}}, \bibinfo {author} {\bibfnamefont {C.}~\bibnamefont {Collins}}, \bibinfo {author} {\bibfnamefont {K.~M.}\ \bibnamefont {Hudek}}, \bibinfo {author} {\bibfnamefont {J.}~\bibnamefont {Mizrahi}}, \bibinfo {author} {\bibfnamefont {J.~D.}\ \bibnamefont {Wong-Campos}}, \bibinfo {author} {\bibfnamefont {S.}~\bibnamefont {Allen}}, \bibinfo {author} {\bibfnamefont {J.}~\bibnamefont {Apisdorf}}, \bibinfo {author} {\bibfnamefont
  {P.}~\bibnamefont {Solomon}}, \bibinfo {author} {\bibfnamefont {M.}~\bibnamefont {Williams}}, \bibinfo {author} {\bibfnamefont {A.~M.}\ \bibnamefont {Ducore}}, \bibinfo {author} {\bibfnamefont {A.}~\bibnamefont {Blinov}}, \bibinfo {author} {\bibfnamefont {S.~M.}\ \bibnamefont {Kreikemeier}}, \bibinfo {author} {\bibfnamefont {V.}~\bibnamefont {Chaplin}}, \bibinfo {author} {\bibfnamefont {M.}~\bibnamefont {Keesan}}, \bibinfo {author} {\bibfnamefont {C.}~\bibnamefont {Monroe}}, \ and\ \bibinfo {author} {\bibfnamefont {J.}~\bibnamefont {Kim}},\ }\bibfield  {title} {\enquote {\bibinfo {title} {Benchmarking an 11-qubit quantum computer},}\ }\href {\doibase 10.1038/s41467-019-13534-2} {\bibfield  {journal} {\bibinfo  {journal} {Nature Communications}\ }\textbf {\bibinfo {volume} {10}},\ \bibinfo {pages} {5464} (\bibinfo {year} {2019})}\BibitemShut {NoStop}%
\bibitem [{\citenamefont {Pogorelov}\ \emph {et~al.}(2021)\citenamefont {Pogorelov}, \citenamefont {Feldker}, \citenamefont {Marciniak}, \citenamefont {Postler}, \citenamefont {Jacob}, \citenamefont {Krieglsteiner}, \citenamefont {Podlesnic}, \citenamefont {Meth}, \citenamefont {Negnevitsky}, \citenamefont {Stadler}, \citenamefont {H\"ofer}, \citenamefont {W\"achter}, \citenamefont {Lakhmanskiy}, \citenamefont {Blatt}, \citenamefont {Schindler},\ and\ \citenamefont {Monz}}]{pogorelov2021compact}%
  \BibitemOpen
  \bibfield  {author} {\bibinfo {author} {\bibfnamefont {I.}~\bibnamefont {Pogorelov}}, \bibinfo {author} {\bibfnamefont {T.}~\bibnamefont {Feldker}}, \bibinfo {author} {\bibfnamefont {Ch.~D.}\ \bibnamefont {Marciniak}}, \bibinfo {author} {\bibfnamefont {L.}~\bibnamefont {Postler}}, \bibinfo {author} {\bibfnamefont {G.}~\bibnamefont {Jacob}}, \bibinfo {author} {\bibfnamefont {O.}~\bibnamefont {Krieglsteiner}}, \bibinfo {author} {\bibfnamefont {V.}~\bibnamefont {Podlesnic}}, \bibinfo {author} {\bibfnamefont {M.}~\bibnamefont {Meth}}, \bibinfo {author} {\bibfnamefont {V.}~\bibnamefont {Negnevitsky}}, \bibinfo {author} {\bibfnamefont {M.}~\bibnamefont {Stadler}}, \bibinfo {author} {\bibfnamefont {B.}~\bibnamefont {H\"ofer}}, \bibinfo {author} {\bibfnamefont {C.}~\bibnamefont {W\"achter}}, \bibinfo {author} {\bibfnamefont {K.}~\bibnamefont {Lakhmanskiy}}, \bibinfo {author} {\bibfnamefont {R.}~\bibnamefont {Blatt}}, \bibinfo {author} {\bibfnamefont {P.}~\bibnamefont {Schindler}}, \ and\ \bibinfo {author}
  {\bibfnamefont {T.}~\bibnamefont {Monz}},\ }\bibfield  {title} {\enquote {\bibinfo {title} {Compact ion-trap quantum computing demonstrator},}\ }\href {\doibase 10.1103/PRXQuantum.2.020343} {\bibfield  {journal} {\bibinfo  {journal} {PRX Quantum}\ }\textbf {\bibinfo {volume} {2}},\ \bibinfo {pages} {020343} (\bibinfo {year} {2021})}\BibitemShut {NoStop}%
\bibitem [{\citenamefont {Piltz}\ \emph {et~al.}(2016)\citenamefont {Piltz}, \citenamefont {Sriarunothai}, \citenamefont {Ivanov}, \citenamefont {Wölk},\ and\ \citenamefont {Wunderlich}}]{piltz2016varsatile}%
  \BibitemOpen
  \bibfield  {author} {\bibinfo {author} {\bibfnamefont {Christian}\ \bibnamefont {Piltz}}, \bibinfo {author} {\bibfnamefont {Theeraphot}\ \bibnamefont {Sriarunothai}}, \bibinfo {author} {\bibfnamefont {Svetoslav~S.}\ \bibnamefont {Ivanov}}, \bibinfo {author} {\bibfnamefont {Sabine}\ \bibnamefont {Wölk}}, \ and\ \bibinfo {author} {\bibfnamefont {Christof}\ \bibnamefont {Wunderlich}},\ }\bibfield  {title} {\enquote {\bibinfo {title} {Versatile microwave-driven trapped ion spin system for quantum information processing},}\ }\href {https://www.science.org/doi/abs/10.1126/sciadv.1600093} {\bibfield  {journal} {\bibinfo  {journal} {Science Advances}\ }\textbf {\bibinfo {volume} {2}},\ \bibinfo {pages} {e1600093} (\bibinfo {year} {2016})}\BibitemShut {NoStop}%
\bibitem [{\citenamefont {Hahn}\ \emph {et~al.}(2019)\citenamefont {Hahn}, \citenamefont {Zarantonello}, \citenamefont {Schulte}, \citenamefont {Bautista-Salvador}, \citenamefont {Hammerer},\ and\ \citenamefont {Ospelkaus}}]{Hahn2019integrated}%
  \BibitemOpen
  \bibfield  {author} {\bibinfo {author} {\bibfnamefont {H.}~\bibnamefont {Hahn}}, \bibinfo {author} {\bibfnamefont {G.}~\bibnamefont {Zarantonello}}, \bibinfo {author} {\bibfnamefont {M.}~\bibnamefont {Schulte}}, \bibinfo {author} {\bibfnamefont {A.}~\bibnamefont {Bautista-Salvador}}, \bibinfo {author} {\bibfnamefont {K.}~\bibnamefont {Hammerer}}, \ and\ \bibinfo {author} {\bibfnamefont {C.}~\bibnamefont {Ospelkaus}},\ }\bibfield  {title} {\enquote {\bibinfo {title} {Integrated 9be+ multi-qubit gate device for the ion-trap quantum computer},}\ }\href {\doibase 10.1038/s41534-019-0184-5} {\bibfield  {journal} {\bibinfo  {journal} {npj Quantum Information}\ }\textbf {\bibinfo {volume} {5}},\ \bibinfo {pages} {70} (\bibinfo {year} {2019})}\BibitemShut {NoStop}%
\bibitem [{\citenamefont {Fishman}\ \emph {et~al.}(2008)\citenamefont {Fishman}, \citenamefont {De~Chiara}, \citenamefont {Calarco},\ and\ \citenamefont {Morigi}}]{fishman2008structural}%
  \BibitemOpen
  \bibfield  {author} {\bibinfo {author} {\bibfnamefont {Shmuel}\ \bibnamefont {Fishman}}, \bibinfo {author} {\bibfnamefont {Gabriele}\ \bibnamefont {De~Chiara}}, \bibinfo {author} {\bibfnamefont {Tommaso}\ \bibnamefont {Calarco}}, \ and\ \bibinfo {author} {\bibfnamefont {Giovanna}\ \bibnamefont {Morigi}},\ }\bibfield  {title} {\enquote {\bibinfo {title} {Structural phase transitions in low-dimensional ion crystals},}\ }\href {\doibase 10.1103/PhysRevB.77.064111} {\bibfield  {journal} {\bibinfo  {journal} {Phys. Rev. B}\ }\textbf {\bibinfo {volume} {77}},\ \bibinfo {pages} {064111} (\bibinfo {year} {2008})}\BibitemShut {NoStop}%
\bibitem [{\citenamefont {Itano}\ \emph {et~al.}(1995)\citenamefont {Itano}, \citenamefont {Bergquist}, \citenamefont {Bollinger},\ and\ \citenamefont {Wineland}}]{wayne1995cooling}%
  \BibitemOpen
  \bibfield  {author} {\bibinfo {author} {\bibfnamefont {Wayne~M}\ \bibnamefont {Itano}}, \bibinfo {author} {\bibfnamefont {J~C}\ \bibnamefont {Bergquist}}, \bibinfo {author} {\bibfnamefont {J~J}\ \bibnamefont {Bollinger}}, \ and\ \bibinfo {author} {\bibfnamefont {D~J}\ \bibnamefont {Wineland}},\ }\bibfield  {title} {\enquote {\bibinfo {title} {Cooling methods in ion traps},}\ }\href {\doibase 10.1088/0031-8949/1995/T59/013} {\bibfield  {journal} {\bibinfo  {journal} {Physica Scripta}\ }\textbf {\bibinfo {volume} {1995}},\ \bibinfo {pages} {106} (\bibinfo {year} {1995})}\BibitemShut {NoStop}%
\bibitem [{\citenamefont {Mao}\ \emph {et~al.}(2021)\citenamefont {Mao}, \citenamefont {Xu}, \citenamefont {Mei}, \citenamefont {Zhao}, \citenamefont {Jiang}, \citenamefont {Wang}, \citenamefont {Chang}, \citenamefont {He}, \citenamefont {Yao}, \citenamefont {Zhou}, \citenamefont {Wu},\ and\ \citenamefont {Duan}}]{mao2021experimental}%
  \BibitemOpen
  \bibfield  {author} {\bibinfo {author} {\bibfnamefont {Z.-C.}\ \bibnamefont {Mao}}, \bibinfo {author} {\bibfnamefont {Y.-Z.}\ \bibnamefont {Xu}}, \bibinfo {author} {\bibfnamefont {Q.-X.}\ \bibnamefont {Mei}}, \bibinfo {author} {\bibfnamefont {W.-D.}\ \bibnamefont {Zhao}}, \bibinfo {author} {\bibfnamefont {Y.}~\bibnamefont {Jiang}}, \bibinfo {author} {\bibfnamefont {Y.}~\bibnamefont {Wang}}, \bibinfo {author} {\bibfnamefont {X.-Y.}\ \bibnamefont {Chang}}, \bibinfo {author} {\bibfnamefont {L.}~\bibnamefont {He}}, \bibinfo {author} {\bibfnamefont {L.}~\bibnamefont {Yao}}, \bibinfo {author} {\bibfnamefont {Z.-C.}\ \bibnamefont {Zhou}}, \bibinfo {author} {\bibfnamefont {Y.-K.}\ \bibnamefont {Wu}}, \ and\ \bibinfo {author} {\bibfnamefont {L.-M.}\ \bibnamefont {Duan}},\ }\bibfield  {title} {\enquote {\bibinfo {title} {Experimental realization of multi-ion sympathetic cooling on a trapped ion crystal},}\ }\href {\doibase 10.1103/PhysRevLett.127.143201} {\bibfield  {journal} {\bibinfo  {journal} {Phys. Rev. Lett.}\
  }\textbf {\bibinfo {volume} {127}},\ \bibinfo {pages} {143201} (\bibinfo {year} {2021})}\BibitemShut {NoStop}%
\bibitem [{\citenamefont {Barrett}\ \emph {et~al.}(2003)\citenamefont {Barrett}, \citenamefont {DeMarco}, \citenamefont {Schaetz}, \citenamefont {Meyer}, \citenamefont {Leibfried}, \citenamefont {Britton}, \citenamefont {Chiaverini}, \citenamefont {Itano}, \citenamefont {Jelenkovi\ifmmode~\acute{c}\else \'{c}\fi{}}, \citenamefont {Jost}, \citenamefont {Langer}, \citenamefont {Rosenband},\ and\ \citenamefont {Wineland}}]{barrett2003sympathetic}%
  \BibitemOpen
  \bibfield  {author} {\bibinfo {author} {\bibfnamefont {M.~D.}\ \bibnamefont {Barrett}}, \bibinfo {author} {\bibfnamefont {B.}~\bibnamefont {DeMarco}}, \bibinfo {author} {\bibfnamefont {T.}~\bibnamefont {Schaetz}}, \bibinfo {author} {\bibfnamefont {V.}~\bibnamefont {Meyer}}, \bibinfo {author} {\bibfnamefont {D.}~\bibnamefont {Leibfried}}, \bibinfo {author} {\bibfnamefont {J.}~\bibnamefont {Britton}}, \bibinfo {author} {\bibfnamefont {J.}~\bibnamefont {Chiaverini}}, \bibinfo {author} {\bibfnamefont {W.~M.}\ \bibnamefont {Itano}}, \bibinfo {author} {\bibfnamefont {B.}~\bibnamefont {Jelenkovi\ifmmode~\acute{c}\else \'{c}\fi{}}}, \bibinfo {author} {\bibfnamefont {J.~D.}\ \bibnamefont {Jost}}, \bibinfo {author} {\bibfnamefont {C.}~\bibnamefont {Langer}}, \bibinfo {author} {\bibfnamefont {T.}~\bibnamefont {Rosenband}}, \ and\ \bibinfo {author} {\bibfnamefont {D.~J.}\ \bibnamefont {Wineland}},\ }\bibfield  {title} {\enquote {\bibinfo {title} {Sympathetic cooling of ${}^{9}{\mathrm{be}}^{+}$ and
  ${}^{24}{\mathrm{mg}}^{+}$ for quantum logic},}\ }\href {\doibase 10.1103/PhysRevA.68.042302} {\bibfield  {journal} {\bibinfo  {journal} {Phys. Rev. A}\ }\textbf {\bibinfo {volume} {68}},\ \bibinfo {pages} {042302} (\bibinfo {year} {2003})}\BibitemShut {NoStop}%
\bibitem [{\citenamefont {Wang}\ \emph {et~al.}(2021)\citenamefont {Wang}, \citenamefont {Luan}, \citenamefont {Qiao}, \citenamefont {Um}, \citenamefont {Zhang}, \citenamefont {Wang}, \citenamefont {Yuan}, \citenamefont {Gu}, \citenamefont {Zhang},\ and\ \citenamefont {Kim}}]{Wang2021single}%
  \BibitemOpen
  \bibfield  {author} {\bibinfo {author} {\bibfnamefont {Pengfei}\ \bibnamefont {Wang}}, \bibinfo {author} {\bibfnamefont {Chun-Yang}\ \bibnamefont {Luan}}, \bibinfo {author} {\bibfnamefont {Mu}~\bibnamefont {Qiao}}, \bibinfo {author} {\bibfnamefont {Mark}\ \bibnamefont {Um}}, \bibinfo {author} {\bibfnamefont {Junhua}\ \bibnamefont {Zhang}}, \bibinfo {author} {\bibfnamefont {Ye}~\bibnamefont {Wang}}, \bibinfo {author} {\bibfnamefont {Xiao}\ \bibnamefont {Yuan}}, \bibinfo {author} {\bibfnamefont {Mile}\ \bibnamefont {Gu}}, \bibinfo {author} {\bibfnamefont {Jingning}\ \bibnamefont {Zhang}}, \ and\ \bibinfo {author} {\bibfnamefont {Kihwan}\ \bibnamefont {Kim}},\ }\bibfield  {title} {\enquote {\bibinfo {title} {Single ion qubit with estimated coherence time exceeding one hour},}\ }\href {\doibase 10.1038/s41467-020-20330-w} {\bibfield  {journal} {\bibinfo  {journal} {Nature Communications}\ }\textbf {\bibinfo {volume} {12}},\ \bibinfo {pages} {233} (\bibinfo {year} {2021})}\BibitemShut {NoStop}%
\bibitem [{\citenamefont {Splatt}\ \emph {et~al.}(2009)\citenamefont {Splatt}, \citenamefont {Harlander}, \citenamefont {Brownnutt}, \citenamefont {Zähringer}, \citenamefont {Blatt},\ and\ \citenamefont {Hänsel}}]{splatt2009deterministic}%
  \BibitemOpen
  \bibfield  {author} {\bibinfo {author} {\bibfnamefont {F}~\bibnamefont {Splatt}}, \bibinfo {author} {\bibfnamefont {M}~\bibnamefont {Harlander}}, \bibinfo {author} {\bibfnamefont {M}~\bibnamefont {Brownnutt}}, \bibinfo {author} {\bibfnamefont {F}~\bibnamefont {Zähringer}}, \bibinfo {author} {\bibfnamefont {R}~\bibnamefont {Blatt}}, \ and\ \bibinfo {author} {\bibfnamefont {W}~\bibnamefont {Hänsel}},\ }\bibfield  {title} {\enquote {\bibinfo {title} {Deterministic reordering of 40ca+ ions in a linear segmented paul trap},}\ }\href {\doibase 10.1088/1367-2630/11/10/103008} {\bibfield  {journal} {\bibinfo  {journal} {New Journal of Physics}\ }\textbf {\bibinfo {volume} {11}},\ \bibinfo {pages} {103008} (\bibinfo {year} {2009})}\BibitemShut {NoStop}%
\bibitem [{\citenamefont {Farhi}\ \emph {et~al.}(2014)\citenamefont {Farhi}, \citenamefont {Goldstone},\ and\ \citenamefont {Gutmann}}]{farhi2014quantum}%
  \BibitemOpen
  \bibfield  {author} {\bibinfo {author} {\bibfnamefont {Edward}\ \bibnamefont {Farhi}}, \bibinfo {author} {\bibfnamefont {Jeffrey}\ \bibnamefont {Goldstone}}, \ and\ \bibinfo {author} {\bibfnamefont {Sam}\ \bibnamefont {Gutmann}},\ }\href@noop {} {\enquote {\bibinfo {title} {A quantum approximate optimization algorithm},}\ } (\bibinfo {year} {2014}),\ \Eprint {http://arxiv.org/abs/1411.4028} {arXiv:1411.4028 [quant-ph]} \BibitemShut {NoStop}%
\bibitem [{\citenamefont {Preskill}(2018)}]{preskill2018quantum}%
  \BibitemOpen
  \bibfield  {author} {\bibinfo {author} {\bibfnamefont {John}\ \bibnamefont {Preskill}},\ }\bibfield  {title} {\enquote {\bibinfo {title} {Quantum computing in the nisq era and beyond},}\ }\href {\doibase https://doi.org/10.22331/q-2018-08-06-79} {\bibfield  {journal} {\bibinfo  {journal} {Quantum}\ }\textbf {\bibinfo {volume} {2}},\ \bibinfo {pages} {79} (\bibinfo {year} {2018})}\BibitemShut {NoStop}%
\bibitem [{\citenamefont {Hadfield}\ \emph {et~al.}(2019)\citenamefont {Hadfield}, \citenamefont {Wang}, \citenamefont {O’Gorman}, \citenamefont {Rieffel}, \citenamefont {Venturelli},\ and\ \citenamefont {Biswas}}]{Hadfield2019}%
  \BibitemOpen
  \bibfield  {author} {\bibinfo {author} {\bibfnamefont {Stuart}\ \bibnamefont {Hadfield}}, \bibinfo {author} {\bibfnamefont {Zhihui}\ \bibnamefont {Wang}}, \bibinfo {author} {\bibfnamefont {Bryan}\ \bibnamefont {O’Gorman}}, \bibinfo {author} {\bibfnamefont {Eleanor~G.}\ \bibnamefont {Rieffel}}, \bibinfo {author} {\bibfnamefont {Davide}\ \bibnamefont {Venturelli}}, \ and\ \bibinfo {author} {\bibfnamefont {Rupak}\ \bibnamefont {Biswas}},\ }\bibfield  {title} {\enquote {\bibinfo {title} {From the quantum approximate optimization algorithm to a quantum alternating operator ansatz},}\ }\href {\doibase 10.3390/a12020034} {\bibfield  {journal} {\bibinfo  {journal} {Algorithms}\ }\textbf {\bibinfo {volume} {12}},\ \bibinfo {pages} {34} (\bibinfo {year} {2019})}\BibitemShut {NoStop}%
\bibitem [{\citenamefont {Blekos}\ \emph {et~al.}(2023)\citenamefont {Blekos}, \citenamefont {Brand}, \citenamefont {Ceschini}, \citenamefont {Chou}, \citenamefont {Li}, \citenamefont {Pandya},\ and\ \citenamefont {Summer}}]{blekos2023review}%
  \BibitemOpen
  \bibfield  {author} {\bibinfo {author} {\bibfnamefont {Kostas}\ \bibnamefont {Blekos}}, \bibinfo {author} {\bibfnamefont {Dean}\ \bibnamefont {Brand}}, \bibinfo {author} {\bibfnamefont {Andrea}\ \bibnamefont {Ceschini}}, \bibinfo {author} {\bibfnamefont {Chiao-Hui}\ \bibnamefont {Chou}}, \bibinfo {author} {\bibfnamefont {Rui-Hao}\ \bibnamefont {Li}}, \bibinfo {author} {\bibfnamefont {Komal}\ \bibnamefont {Pandya}}, \ and\ \bibinfo {author} {\bibfnamefont {Alessandro}\ \bibnamefont {Summer}},\ }\href@noop {} {\enquote {\bibinfo {title} {A review on quantum approximate optimization algorithm and its variants},}\ } (\bibinfo {year} {2023}),\ \Eprint {http://arxiv.org/abs/2306.09198} {arXiv:2306.09198 [quant-ph]} \BibitemShut {NoStop}%
\bibitem [{\citenamefont {Shor}(1999)}]{shor1999polynomial}%
  \BibitemOpen
  \bibfield  {author} {\bibinfo {author} {\bibfnamefont {Peter~W.}\ \bibnamefont {Shor}},\ }\bibfield  {title} {\enquote {\bibinfo {title} {Polynomial-time algorithms for prime factorization and discrete logarithms on a quantum computer},}\ }\href {\doibase 10.1137/S0036144598347011} {\bibfield  {journal} {\bibinfo  {journal} {SIAM Review}\ }\textbf {\bibinfo {volume} {41}},\ \bibinfo {pages} {303--332} (\bibinfo {year} {1999})}\BibitemShut {NoStop}%
\bibitem [{\citenamefont {DeCross}\ \emph {et~al.}(2024)\citenamefont {DeCross}, \citenamefont {Haghshenas}, \citenamefont {Liu}, \citenamefont {Alexeev}, \citenamefont {Baldwin}, \citenamefont {Bartolotta}, \citenamefont {Bohn}, \citenamefont {Chertkov}, \citenamefont {Colina}, \citenamefont {DelVento} \emph {et~al.}}]{decross2024computational}%
  \BibitemOpen
  \bibfield  {author} {\bibinfo {author} {\bibfnamefont {Matthew}\ \bibnamefont {DeCross}}, \bibinfo {author} {\bibfnamefont {Reza}\ \bibnamefont {Haghshenas}}, \bibinfo {author} {\bibfnamefont {Minzhao}\ \bibnamefont {Liu}}, \bibinfo {author} {\bibfnamefont {Yuri}\ \bibnamefont {Alexeev}}, \bibinfo {author} {\bibfnamefont {Charles~H}\ \bibnamefont {Baldwin}}, \bibinfo {author} {\bibfnamefont {John~P}\ \bibnamefont {Bartolotta}}, \bibinfo {author} {\bibfnamefont {Matthew}\ \bibnamefont {Bohn}}, \bibinfo {author} {\bibfnamefont {Eli}\ \bibnamefont {Chertkov}}, \bibinfo {author} {\bibfnamefont {Jonhas}\ \bibnamefont {Colina}}, \bibinfo {author} {\bibfnamefont {Davide}\ \bibnamefont {DelVento}},  \emph {et~al.},\ }\bibfield  {title} {\enquote {\bibinfo {title} {The computational power of random quantum circuits in arbitrary geometries},}\ }\href {https://arxiv.org/abs/2406.02501} {\bibfield  {journal} {\bibinfo  {journal} {arXiv preprint arXiv:2406.02501}\ } (\bibinfo {year} {2024})}\BibitemShut {NoStop}%
\bibitem [{\citenamefont {Quantinuum}(2024)}]{racetrack2024sheet}%
  \BibitemOpen
  \bibfield  {author} {\bibinfo {author} {\bibnamefont {Quantinuum}},\ }\href {https://assets.website-files.com/62b9d45fb3f64842a96c9686/665f5f21ce5b0435bcd32917_Quantinuum%20H2%20Product%20Data%20Sheet%20v1.4%204Jun24.pdf} {\emph {\bibinfo {title} {Quantinuum System Model H2 - Product Data Sheet}}} (\bibinfo {year} {2024})\BibitemShut {NoStop}%
\end{thebibliography}%

\end{document}